\documentclass[aps,twocolumn,prb]{revtex4-1}
\usepackage[version=3]{mhchem} 
\usepackage{graphicx}
\usepackage{dcolumn}
\usepackage{bm}
\usepackage{comment}
\usepackage{epstopdf}
\usepackage{amsmath}
\usepackage{hyperref}
\providecommand{\e}[1]{\ensuremath{\times 10^{#1}}}
\newcommand{\myfrac}[3][0pt]{\genfrac{}{}{}{}{\raisebox{0pt}{$#2$}}{\raisebox{-#1}{$#3$}}}
\begin{document}
\author{A.~C. McRae, G. Wei, and A.~R. Champagne}
\email{a.champagne@concordia.ca}
\affiliation{Department of Physics, Concordia University, Montr\'{e}al, Qu\'{e}bec, H4B 1R6, Canada}
\title[\texttt{achemso}]{Graphene Quantum Strain Transistors}
\date{\today}

\begin{abstract}
There is a wide range of science and applications accessible via the strain engineering of quantum transport in 2D materials. We propose a realistic experimental platform for uniaxial strain engineering of ballistic charge transport in graphene. We then develop an applied theoretical model, based on this platform, to calculate charge conductivity and demonstrate graphene quantum strain transistors (GQSTs). We define GQSTs as mechanically strained ballistic graphene transistors with on/off conductivity ratios $> 10^{4}$, and which can be operated via modest gate voltages. Such devices would permit excellent transistor operations in pristine graphene, where there is no band gap. We consider all dominant uniaxial strain effects on conductivity, while including experimental considerations to guide the realization of the proposal. We predict multiple strain-tunable transport signatures, and demonstrate that a broad range of realistic device parameters lead to robust GQSTs. These devices could find applications in flexible electronic transistors, strain sensors, and valleytronics.
\end{abstract}
\maketitle

\section{Introduction}
High on/off ratio transistors based on pristine bulk graphene have been sought after for a long time, and were previously proposed in idealized ballistic devices under uniaxial strain \cite{Fogler08,Pellegrino11,Cao12}. We name these proposed devices graphene quantum strain transistors (GQSTs). Their conductivity can be turned off, not due to a band gap, but because uniaxial strain can tailor the energy, momentum, and quantum transmission of graphene's ballistic electrons \cite{CastroNeto09}. GQSTs are technologically relevant given that ballistic transport in graphene can persist over lengths up to a micron at room temperature \cite{Mayorov11,Banzerus16}. A state-of-the-art transistor behavior in graphene would be a paradigm shift for flexible electronic devices \cite{Akinwande14}, and the quantum engineering of 2D heterostructures \cite{Iannaccone18}. Unfortunately, a full decade after the first proposal of GQSTs \cite{Fogler08}, there is no clear path forward to realize them experimentally. This bottleneck is widespread amongst many proposals in quantum transport strain engineering (QTSE) in graphene and other 2D materials \cite{Fogler08, Guinea10, Amorim16,Settnes17, Wu17,Naumis17}. We see two main causes for the slow experimental realization of QTSE proposals in 2D materials. Firstly, there is no established experimental platform, with \textit{in situ} tunable mechanical strain, suited for low-temperature quantum transport measurements. Secondly, due to the hypersensitivity of 2D materials to their environment, there are large quantitative discrepancies between the predictions of idealized transport models and experiments. As a concrete example, previous theoretical predictions of the GQST effect were far too optimistic because they treated metal films covering graphene as not affecting its Fermi level\cite{Fogler08}, or omitted important effects of the mechanical strain\cite{Pellegrino11,Cao12,Kitt12,Kitt13} on graphene's Hamiltonian.

Graphene is an ideal system to first bridge the present experiment-theory divide in QTSE of 2D materials. Its extreme mechanical strength, flexibility, and elastic deformation range make graphene's electronics vastly tunable via mechanical deformations \cite{CastroNeto09,Guinea10,Amorim16,Naumis17}, while its defect-free lattice allows ballistic (quantum) transport \cite{Mayorov11,Banzerus16}. It has been shown that a non-uniform strain can add extreme pseudomagnetic fields\cite{Levy10} $\sim$ 100 T to graphene's Dirac Hamiltonian. Mechanical strains can be applied to graphene deposited on flexible\cite{Kim09} or nanoengineered\cite{Zhang18} substrates, as well as in suspended devices\cite{Guan17}. In the short term, uniaxial strain engineering of graphene could demonstrate GQSTs. These would be ideal transistors for flexible electronics due to their high on/off ratio and fast transit frequency \cite{Tan17}, could be used as sensitive strain sensors\cite{Smith16,Chun17}, and permit the development of valley filters for valleytronics \cite{Fujita10,Yesilyurt16}. Beyond uniaxial strain, there are ongoing efforts to use strain-engineering in graphene to explore zero-magnetic-field quantum Hall physics \cite{Levy10,Guinea10,Zhu15, Jiang17, Settnes17} and topological phase transitions \cite{Guassi15,Wu17}.

Here, we first present a powerful experimental platform for uniaxial strain engineering of quantum transport in graphene or other 2D materials. We provide all relevant experimental parameters of both the mechanical strain instrumentation and suspended graphene devices to realize GQSTs. We calculate the mechanically-tunable, thermally-induced, and gate-induced strains in the devices, and predict a widely tunable total mechanical strain ($ 2.6~\% < \varepsilon_{\text{total}} < 5.1~\%$). We show that $\varepsilon_{\text{total}}$ and the Fermi energy, $E_{F}$, in the transistor channel can be independently controlled. We then present a complete model of the ballistic charge conductivity in uniaxially strained graphene, which describes the proposed experimental platform. We include all dominant effects of uniaxial strain, such as the scalar potential\cite{Kretinin13, Naumis17} due to modulation of next-nearest-neighbor (nnn) hopping, and vector potentials\cite{Kitt12,Naumis17} from changes to the nearest-neighbor (nn) hoppings, $\gamma_{n}$, and spacings, $\bm{\delta}_{n}$, where $n=1,2,3$ (see Fig.\ 1). We use experimentally relevant parameters, including a realistic doping level for graphene covered by a metal \cite{Chaves14}, a wide channel $W/L\gg1$ boundary condition \cite{Tworzydlo06}, easy-to-fabricate device dimensions, arbitrary crystal orientations, and achievable strain values.

The main purpose of our detailed calculations is to properly guide experimentalists toward the realization of GQSTs. As such, we detail all needed ingredients to make major quantitative (order of magnitude) improvements in the accuracy of previous models to calculate the strain-dependent conductivity. We report four quantitative transport signatures of the model: a shift of the gate voltage position of the conductivity minimum, a dramatic conductivity suppression, a rich set of Fabry-P\'{e}rot (FP) interferences, and a large modulation of the electron-hole conductivity asymmetry. Finally, we demonstrate that a strong GQST effect ($\sigma_{\text{on/off}}>10^4$) is broadly realistic. The transistors can be turned on/off via a simple gate voltage, while strain is held constant. The simplicity, and detailed description, of the experimental proposal should permit a short-term demonstration of GQSTs. More broadly, we believe that this type of experiment-ready proposal can accelerate the development of QTSE in 2D materials.
\section{Platform for QTSE in 2D Materials}
\begin{figure}
\includegraphics{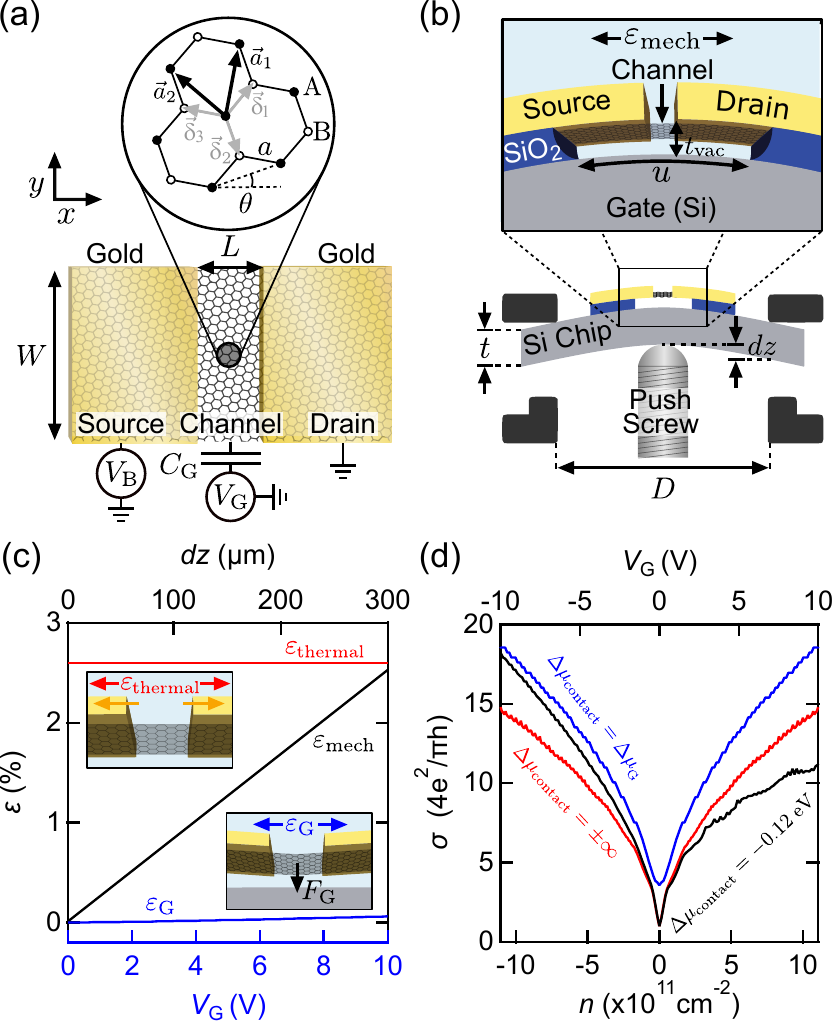}
\caption{Platform for uniaxial quantum transport strain-engineering (QTSE) in graphene. (a) Top-down view of the proposed ballistic graphene transistor geometry. Outset: the graphene lattice, showing the crystal orientation $\theta$ with respect to the $x$-axis. (b) Side views of the proposed graphene device and mechanical strain instrumentation. The mechanical assembly bends the substrate, which strains the suspended channel. (c) The three sources of strain in the channel: the mechanical motion (black, top axis) of the push screw, the thermal contraction (red) at $\sim$ 1 Kelvin, and electrostatic strain (blue, bottom axis) from $V_{\text{G}}$. Insets: visualizing the strain imparted by the thermal contraction of the gold cantilevers (top left), and electrostatic pulling (bottom right). (d) Conductivity versus charge density, $n$ (bottom axis), or $V_{\text{G}}$ (top axis). The data are for an unstrained channel of $L=100$ nm, $W=1000$ nm, with contact doping $\Delta\mu_{\text{contact}}= -0.12$~eV (black), $\pm\infty$ (red), and $\Delta\mu_{\text{contact}}= \Delta\mu_{\text{G}}$ (blue).}
\end{figure}

We propose, in Fig.\ 1, a device geometry and instrumentation to realize graphene quantum strain transistors. This platform combines many important features to connect measurements to a simple theory. The suspended graphene channel is shown in Fig.\ 1(a) -- (b), and its dimensions, $L=$ 100 nm and $W=$ 1000 nm, can be readily fabricated \cite{Yigen14}. The length of the naked graphene channel is defined by gold films directly deposited on top of exfoliated graphene. It was previously reported that such gold films, after annealing, can dope the graphene underneath without disrupting its ballistic transport\cite{Wu12,McRae17}. These gold-covered graphene regions act as contacts to the naked graphene channel. Few-nm sharp $p$--$n$ junctions form between these graphene contacts and the naked channel\cite{Wu12,McRae17}. The large aspect ratio of the channel $W/L \gg$ 1 ensures that quantum transport is not affected by the atomic disorder at the edges of the exfoliated crystal, such that a smooth boundary condition in the $y$-direction is justified \cite{Tworzydlo06}. Additionally, because the channel is mechanically clamped across its entire width and $W/L\gg$ 1, there is no significant scrolling of the free edges \cite{Guan17}. Thus, the channel closely matches an ideal rectangle and can be described with a simple analytic ballistic transport model.

The outset of Fig.\ 1(a) shows the graphene lattice with nearest-neighbor spacing $a=1.42$~\AA, the two-atom basis (A,B), primitive lattice vectors $\bm{a}_{i}$, and nearest-neighbor vectors $\bm{\delta}_{n}$. Also pictured is the crystal orientation, $\theta$, defined as the angle between the $x$-axis of the device and the zig-zag direction of the crystal. This angle can be measured prior to the gold film deposition using polarized Raman spectroscopy \cite{Huang09}, or STM imaging \cite{Andrei12}. Figure 1(b) shows suspended gold beams, $\approx$ 120 nm thick, used to anchor the suspended graphene. The total suspension length of the gold cantilevers and channel is $u =$ 900 nm. This suspension length can be tuned by etching the SiO$_2$ under the gold with a standard anisotropic wet etch \cite{Yigen14}. Upon bending the silicon substrate, the gold cantilevers amplify the mechanical strain applied to the naked graphene channel. The suspension height of the channel above the doped-silicon back-gate is $t_{\text{vac}}=$ 50 nm. The device's suspension enables its \textit{in situ} thermal annealing, so as to reach the ballistic transport regime in both the channel \cite{Bolotin08} and suspended contacts \cite{Tayari15, McRae17}. A standard dc transport circuit (Fig.\ 1(a)) can be used for both the annealing and transport measurements. An important aspect of this device design is the large areas of the two gold mechanical anchors ($\sim\mu$m$^2$) permitting slippage-free clamping of the graphene channel.

The proposed mechanical strain instrumentation shown in the lower portion of Fig.\ 1(b) is very similar to a previously reported mechanical break-junction assembly \cite{Champagne05}. It was shown to operate accurately over the mechanical range needed for the present proposal, and at low-temperatures and in high magnetic fields\cite{Parks07, Parks10}. This instrumentation is better suited for QTSE in 2D materials than recent methods \cite{Goldsche18, Sarwat18} which may not be compatible with low-temperatures, or do not offer an independent control of both strain and charge density in the channel. A finely-polished macroscopic push screw is used to reversibly bend a 200 $\mu$m-thick silicon substrate between two anchoring points, spaced apart by $D=$ 8 mm. The expected channel strain from the motion of the push screw is given by \cite{Champagne05}, $\varepsilon_{\text{mech}}=(3ut/D^2)dz/L$, where $L$ is the naked graphene channel length, $u$ is the total suspended length of the gold cantilevers and channel, $t$ is substrate thickness, $D$ is the distance between the anchoring points, and $dz$ is the vertical displacement of the push screw. Based on previous experiments \cite{Champagne05}, the range of $dz$ is up to $\sim$ 300 $\mu$m, giving a range of mechanical strain $\varepsilon_{\text{mech}} \geq 2.5~\% $. This instrumentation is modular such that the clamping of the substrate and graphene flake could be modified to permit, for instance, triaxial strain engineering \cite{Guinea10,Settnes17}.

Figure 1(c) details the main sources of strain in the suspended channel. Because device nanofabrication is done at room temperature, a substantial amount of strain will be generated when cooling the devices to cryogenic temperatures. The top inset of Fig.\ 1(c) shows the contribution to this thermal strain, $\varepsilon_{\text{thermal}}$, arising from the contraction of the gold cantilevers \cite{Nix41}. As detailed in Supplementary Information S1, we find $\varepsilon_{\text{thermal}}=2.6 \pm 0.1~\%$ for devices near $\sim$ 1 Kelvin. As shown in Fig.\ 1(c), this $\varepsilon_{\text{thermal}}$ is independent of either gate voltage, $V_{\text{G}}$, or mechanical displacement of the push screw, $dz$. At a fixed temperature, this constant thermal strain will not hinder the exploration of strain engineering proposals. A second source of strain in the channel is the stretching $\varepsilon_{\text{G}}$ caused by the electrostatic force\cite{Fogler08} from $V_{\text{G}}$, as shown in the bottom inset of Fig.\ 1(c). Given the very short length of the graphene channel, $L = 100$ nm, and the sturdiness of the gold cantilevers, we calculate that $\varepsilon_{\text{G}}$ is very small ($\sim$ 0.01~$\%$) over the $-$10 V $<V_{\text{G}}<$ 10 V range relevant for experiments. This leaves the mechanical push screw $0 \leq\varepsilon_{\text{mech}} \leq 2.5$~$\%$ (black data) as the only way to tune the total strain, $\varepsilon_{\text{total}}=\varepsilon_{\text{thermal}}+\varepsilon_{\text{G}}+\varepsilon_{\text{mech}} = 2.6$ -- $5.1$~$\%$. The proposed instrumentation permits broad, and independent, control of the uniaxial strain and charge density in the graphene channel.

To complete the GQST proposal, we require devices with adequate ballistic contact regions for the ballistic channel. This will permit the use of a simple analytical ballistic model for both the contacts and channel, and enable high on/off ratio GQSTs. Indeed, as we will see below, the graphene contacts' Fermi energy $\Delta\mu_{\text{contact}}$ must be comparable or smaller than the strain-induced potentials in the channel for optimal GQST operation. The source and drain contacts in our devices are sections of the same graphene crystal as the naked channel, but are located under gold cantilevers (Fig.\ 1(a)--(b)). The very long injection length of electrons from gold into graphene \cite{Sundaram11,McRae17} ensures that the electrons entering the naked graphene channel come from the graphene electrodes, and not directly from the gold film. The $\Delta\mu_{\text{contact}}$ in the graphene contacts arises from the charge transfer due to the work function difference between the metal and graphene\cite{Chaves14}. This contact doping determines how many subbands (ballistic modes) are occupied in the contacts. Each of these contact modes may be transmitted across the channel, depending on $V_{\text{G}}$ and the strain-induced potentials.

We chose gold as the metal covering the graphene contacts, because its work function leads to the needed contact doping \cite{Chaves14,Heinze02} ($|\Delta\mu_{\text{contact}}|= 0.05$ -- $0.25$ eV). The exact $\Delta\mu_{\text{contact}}$ depends on the oxygen content of the gold film, since it modifies the gold work function. The oxygen content can be adjusted \textit{in situ} via Joule annealing \cite{McRae17}. For the purpose of the present work we use a median value\cite{McRae17} of $|\Delta\mu_{\text{contact}}| = 0.12$~eV with \textit{p}-doping. Figure 1(d) shows the calculated ballistic conductivity, $\sigma$, as a function of the charge density in the channel, $n$, or equivalently $V_{\text{G}}$ on the top axis, when $\varepsilon_{\text{total}} = 0$. We based this calculation on the standard ballistic transport model for a $W/L \gg 1$ geometry \cite{Tworzydlo06}. To highlight the impact of contact doping on the $\sigma$ - $V_{\text{G}}$ characteristics, we plot in Fig. 1(d) the conductivity for $\Delta\mu_{\text{contact}} = - $ 0.12 eV (black), $\pm\infty$ (red), and $\Delta\mu_{\text{contact}}= \Delta\mu_{\text{G}}$ (blue). The inclusion of a realistic $\Delta\mu_{\text{contact}}$ becomes critical when strain is applied. An infinite contact doping makes $\sigma$ completely insensitive to strain, while the previously used approximation \cite{Fogler08} of $\Delta\mu_{\text{contact}}= \Delta\mu_{\text{G}}$ grossly exaggerates the strain dependence. We note that our main conclusion that GQSTs are feasible is preserved over a broad range of realistic $\Delta\mu_{\text{contact}}$, as we will show below.
\section{Applied Theory for Uniaxial QTSE in Graphene}
\begin{figure*}
\includegraphics{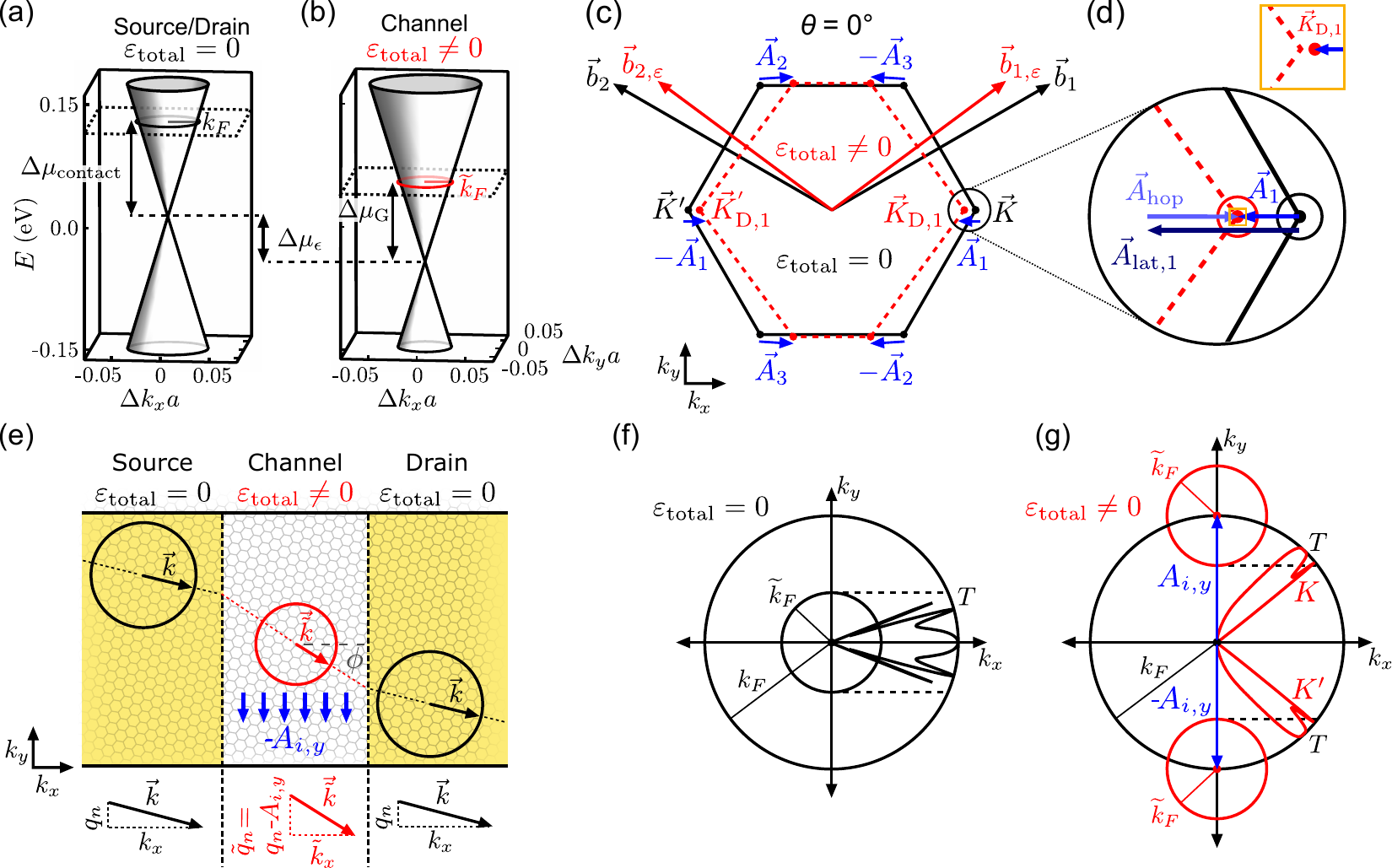}
\caption{Applied theory for uniaxial QTSE in graphene. (a) Dirac cone and Fermi circle in the unstrained source/drain graphene contacts, and (b) in the strained graphene channel. (c) Unstrained (black) and uniaxially strained (red) first Brillouin zone (FBZ) of graphene when $\theta=0^{\circ}$. The strain value in this figure is exaggerated, $\varepsilon_{\text{total}}=20$~$\%$, to make its effects clearly visible. (d) Under strain, the Dirac point shifts define gauge vector potentials (blue arrows), $\bm{A}_{i} = \bm{A}_{\text{lat},i} + \bm{A}_{\text{hop}} $. The outset shows that the corner of the FBZ does not coincide with the Dirac point under strain. (e) Charge carrier wave vectors in the source, channel, and drain of the uniaxially strained ballistic transistor. The ballistic modes are labelled with their $y$-component wave number $q_{n}$. The $A_{i,y}$ in the channel modifies the propagation angle, $\phi$, and transmission probability, $T$, of the carriers. (f) -- (g) $T$ of the conduction modes for $A_{i,y} =0$ and $A_{i,y} = k_F$ respectively. The circles represent the Fermi surfaces in the contacts (big circles) and channel (small circles), while the solid curves show the $T$ (origin $= 0$, outer circle $=1$) versus the incidence angle on the channel (polar axis), when $\theta=15^{\circ}$ and $\Delta\mu_\text{contact}=-$ 0.12 eV. The boundary condition in $y$ imposes transversal momentum conservation (dashed lines), and explains the zero transmission at some angles.}
\end{figure*}
In this section, we translate the above instrumentation and device design into an applied theoretical model to calculate the transmission probability of charge carriers across the device as it is uniaxially strained. We consider all strain effects to first order, given that $\varepsilon_{\text{total}} \ll 1$. The uniaxially-strained-graphene Dirac Hamiltonian, shown in Eq.\ 1, is similar to previously reported ones \cite{Tworzydlo06, Pereira09, Pellegrino11, Cao12,Kitt12, Naumis17}, but includes all relevant experimental considerations (Supplementary Information S2). The details, and hierarchy, of all experimental and theoretical uncertainties in our proposal are discussed in Supplementary Information S3. Given the smooth-edge boundary condition for our wide aspect ratio devices, the Hamiltonians in the $K_{i}$ and $K'_{i}$ valleys are decoupled, $i =$ 1, 2, 3. Equation 1 describes the $K_{i}$ valleys in the channel, while reversing the sign of the $\bm{A}_{i}$ gives the correct expression in the $K'_{i}$ valleys.

\begin{widetext}
\begin{equation}
H_{K_{i}}=\hbar v_{F}(\bar{\bm{I}}+(1-\beta)\bar{\bm{\varepsilon}})\cdot\bm{\sigma}\cdot(\tilde{\bm{k}}-\bm{A}_{i}) + \Delta\mu_{\text{G}} + \Delta\mu_{\varepsilon}
\end{equation}
\end{widetext}
The $\tilde{\bm{k}}$ refers to the carrier wave vector in the strained channel. The pseudospin operator $\bm{\sigma}=(\sigma_x,\sigma_y)$ is represented by the Pauli matrices and acts on the two-component spinor wavefunction referring to the A and B sublattices. The pseudospin orientation is either parallel (up) or anti-parallel (down) with the generalized wave vector $\tilde{\bm{k}} - \bm{A}_{i}$. The matrices $\bar{\bm{I}}$ and $\bar{\bm{\varepsilon}}$ are respectively the identity matrix and strain tensor. In the device's $x$ - $y$ coordinates, $\bar{\bm{\varepsilon}}$ has elements $\varepsilon_{xx}=\varepsilon_{\text{total}}$, $\varepsilon_{yy}=-\nu \varepsilon_{\text{total}}$ and $\varepsilon_{xy}=\varepsilon_{yx}=0$, where $\nu=0.165$ is the Poisson ratio \cite{Naumis17}. The term $\Delta\mu_{\text{G}}=\hbar v_F\sqrt{\pi C_{\text{G}} (V_{\text{G}}-V_{\text{D}}) /e}$ is the gate-induced electrostatic potential in the channel, where $v_{F}=(3/2)\gamma_0 a/\hbar=8.8\e{5}$  m/s is the Fermi velocity, $\gamma_0 = -2.7$~eV, and $C_{\text{G}} =$ 1.7 $\times 10^{-8}$ F/cm$^2$ is the gate-channel capacitance.

The Hamiltonian in the source/drain graphene contacts is similar to Eq.\ 1, but the carrier wave vector $\tilde{\bm{k}}$ is replaced by $\bm{k}$ for clarity, the strain-induced terms are set to zero, and $\Delta\mu_{\text{G}}$ is replaced with $\Delta\mu_{\text{contact}}$. As per Eq.\ 1, uniaxial strain has three main qualitative effects on the channel's band structure: a downward shift of the Fermi energy which can be described by a scalar potential $\Delta\mu_{\varepsilon}$, crystal momentum shifts of the Dirac points which can be described by gauge vector potentials $\bm{A}_{i}$, and an anisotropic warping the Dirac cones which corresponds to a direction-dependent Fermi velocity $\bar{\bm{v}}_{F} =  v_{F}(\bar{\bm{I}}+(1-\beta)\bar{\bm{\varepsilon}})$. For uniaxial strain, $\bar{\bm{v}}_{F}$ has only diagonal elements $v_{F,xx}=1+(1-\beta)\varepsilon_{\text{total}}$ and $v_{F,yy}=1-(1-\beta)\nu\varepsilon_{\text{total}}$. The parameter $\beta=- d(\ln \gamma_{0})/d(\ln a) \approx 2.5$ is the electronic Gr\"{u}neisen parameter \cite{Naumis17}. We discuss in Fig.\ 2 the origin and magnitude of the strain-induced scalar and vector potentials, and how they impact charge transport. We then calculate the transmission probability of individual conduction modes, and sum them to obtain the device's charge conductivity.

Figure 2(a) -- (b) shows the low energy dispersion and Fermi circle around a Dirac point in, respectively, the source/drain contacts and the strained channel. The magnitude of the Fermi wave vector in Fig.\ 2(a) is set by the contact doping, $k_F=\Delta\mu_{\text{contact}}/\hbar v_F$. In the strained channel, as shown in Fig.\ 2(b), $\tilde{k}_F$ depends on both the strain-induced scalar potential and the gate-induced doping, $\tilde{k}_{F}=(\Delta\mu_{\text{G}}+\Delta\mu_{\varepsilon})/\hbar v_{F}$. The term $\Delta\mu_{\varepsilon}$ arises from the strain-dependence of the nnn hopping\cite{Kretinin13} (Supplementary Information S2), $\gamma^{'}_{0} \approx -0.3$~eV. The nnn hopping does not connect lattice sites A and B, and is thus not related to the pseudospin of charge carriers. The consequence of this scalar potential is a shift of the low energy band structure, given by $\Delta\mu_{\varepsilon}=g_{\varepsilon}(1-\nu)\varepsilon_{\text{total}}$, where\cite{Choi10} $g_{\varepsilon}\approx 3.0$~eV.

The strain-induced changes to the nearest-neighbor hopping\cite{Kitt12} $\gamma_0$ are incorporated in the Dirac Hamiltonian of graphene as gauge vector potentials $\bm{A}_{i}$. Figure 2(c), shows the unstrained (black) and strained (red) first Brillouin zones (FBZs) for a strain along the zig-zag direction ($\theta=0^{\circ}$). To better visualize the $\bm{A}_{i}$ in Fig.\ 2(c) -- (d), we show $\varepsilon_{\text{total}}=$ 20~$\%$, but restate that we only studied $\varepsilon_{\text{total}}\leq 5.1$~$\%$. To understand conceptually the origin of the $\bm{A}_{i}$, we need to go back to the tight-binding Hamiltonian of graphene\cite{Kitt12, Naumis17} which led to Eq.\ 1, $H_{0}= \gamma_{0}\sum_{\tilde{k}}\sum_{n=1}^{3}(\exp[-i\tilde{\bm{k}}\cdot\bm{\delta}_{n}]a_{\tilde{k}}^{\dag}b_{\tilde{k}} + \text{ H. c.})$. To first order in strain, there are two main modifications to this tight-binding Hamiltonian. Firstly, the stretching of the nn distances, $\bm{\delta}_{n}$, modifies the Hamiltonian with $\bm{\delta}_{n}\rightarrow \left(\bm{\bar{I}}+\bm{\bar{\varepsilon}}\right)\cdot\bm{\delta}_{n}$. Secondly, the value of the nn hopping is modified as $\gamma_{0}\rightarrow \gamma_{0}\exp\left[-\beta \left\{ \left(\left|\left(\bm{\bar{I}}+\bm{\bar{\varepsilon}}\right)\cdot\bm{\delta}_{n}\right| /a\right) - 1\right\}\right]$. Both of these effects can be rewritten as gauge vector potentials\cite{Kitt12} in Eq.\ 1, respectively as $\bm{A}_{\text{lat},i} = -\bm{\bar{\varepsilon}}\cdot \bm{K}_{i}$, and $\bm{A}_{\text{hop}} = \beta/2a (\varepsilon_{xx}-\varepsilon_{yy},-2\varepsilon_{xy})$. The total gauge potentials are then $\bm{A}_{i}=\bm{A}_{\text{lat},i}+\bm{A}_{\text{hop}}$. The signs of both components of the vector potentials are opposite in the $K_{i}$ and $K'_{i}$ valleys. To make this model applicable to real devices, we generalize (Supplementary Information S2) the vector potentials to an arbitrary crystal orientation $\theta$ with respect to the device's $x$-axis \cite{Kitt12}. We find,
\begin{subequations}
\begin{eqnarray}
&\bm{A}_{i}&=\bm{A}_{\text{lat},i}+\bm{A}_{\text{hop}}
  \\[10pt]
&\bm{A}_{\text{hop}}&=\frac{\beta\varepsilon(1+\nu)}{2a} \begin{pmatrix}
 \cos3\theta\\ \sin3\theta \end{pmatrix}
  \\[10pt]
&\bm{A}_{\text{lat,1}}&=\frac{4\pi\varepsilon}{3\sqrt{3}a} \begin{pmatrix}
 -\cos\theta\\ \nu\sin\theta \end{pmatrix}
 \\[10pt]
&\bm{A}_{\text{lat,2}}&=\frac{2\pi\varepsilon}{3a} \begin{pmatrix}
 \frac{1}{\sqrt{3}}\cos\theta +\sin\theta \\ -\frac{1}{\sqrt{3}}\nu\sin\theta +\nu\cos\theta \end{pmatrix}
 \\[10pt]
&\bm{A}_{\text{lat,3}}&=\frac{2\pi\varepsilon}{3a} \begin{pmatrix}
 \frac{1}{\sqrt{3}}\cos\theta -\sin\theta \\ -\frac{1}{\sqrt{3}}\nu\sin\theta -\nu\cos\theta \end{pmatrix}
\end{eqnarray}
\end{subequations}

Because of lattice symmetry, $0^{\circ}<\theta<60^{\circ}$. The vector potentials from Eq.\ 2 are plotted versus $\theta$ in Supplementary Information S2. An illustration of the vector potentials from Eq.\ 2, at $\theta=0^{\circ}$, is shown in Fig.\ 2(c). The $\bm{A}_{i}$ represent the displacements from the original unstrained Dirac points to the strained ones. Figure 2(d) details the contributions from the terms $\bm{A}_{\text{lat},i}$ and $\bm{A}_{\text{hop}}$ to the new Dirac point locations $\bm{K}_{\text{D},i}= \bm{K}_{i} + \bm{A}_{\text{lat},i} + \bm{A}_{\text{hop}}$. As shown in the outset of Fig.\ 2(d), the new Dirac points no longer coincide with the corners of the FBZ\cite{Pereira09}. We remark that the two terms in the vector potential are of comparable magnitude, and both significantly affect the Hamiltonian and transport. However, in most previous work, the focus has been on studying pseudomagnetic fields\cite{Naumis17}, $\bm{B}_{\text{p}} = \nabla\times\bm{A}$. Because $\bm{A}_{\text{lat},i}$ always has zero curl \cite{Kitt13}, it has often been neglected. It is however crucial to include it in our study, which is focused on the direct consequences of the gauge vector potentials as in an Aharonov-Bohm experiment.

To clarify the connection between Eq.\ 1 and ballistic conductivity in our devices, we show in Fig.\ 2(e) an example of a charge carrier trajectory as it moves from the source (unstrained) to the channel (strained), and then the drain electrode (unstrained). The trajectory is described by the momentum wave vector in the source, ${\bm{k}} = k_x \hat{x} + q_n \hat{y}$, and in the channel, ${\bm{\tilde{k}}} = \tilde{k}_{x} \hat{x} + \tilde{q}_n \hat{y}$. The transverse boundary condition conserves the $y$-momentum throughout the device\cite{Fogler08}, such that $\tilde{q}_n=q_n-A_{i,y}$. This strain-induced shift in $y$-momentum alters the propagation angle, $\phi$, of the carriers, and their Klein transmission at the strained/unstrained interfaces. Using Eq.\ 1 and matching the carrier's wavefunction at the potential steps along $x$, we solve for the transmission probability $T_{\xi,i,n}$ for the conduction mode $n$, in valley $\xi,i$, where $\xi= \pm 1$ for the $K_{i}$ and $K'_{i}$ valleys, respectively, and $i = 1,2,3$. We find
\begin{widetext}
\begin{equation}
T_{\xi,i,n}=\myfrac[4pt]{(\frac{v_{F,xx}}{v_F}k_x\tilde{k}_{x})^2}{(\frac{v_{F,xx}}{v_F}k_x\tilde{k}_{x})^2\cos^2 [\tilde{k}_{x}L]+(k_{F}\tilde{k}_{F}-\frac{v_{F,yy}}{v_F} q_n(q_n-\xi A_{i,y}))^2 \sin^2 [\tilde{k}_{x}L]},
\end{equation}
\end{widetext}
where $q_n=\frac{\pi}{W}(n+\tfrac{1}{2})$ is the quantized transversal momentum for the mode $n$, $k_x=({k_{F}^2-q_{n}^2})^{1/2}$, and $\tilde{k}_{x}=v_{F,xx}^{-1}[{v_F^2 \tilde{k}_F^2-v_{F,yy}^2(q_{n}-\xi A_{i,y})^2}]^{1/2}$. One major insight visible in Eq.\ 3 is that only the $y$ components of the ${\bm{A}}_i$ affect transmission. From Eq.\ 2 we see the magnitudes of the $A_{i,y}$ are maximized when $\theta = 30^{{\circ}}$, \textit{i.e.} when the strain is along the armchair edge (Supplementary Information S2). We calculate the charge conductivity of the device by properly summing the contributions from all relevant modes:
\begin{equation}
\sigma=\frac{L}{W}\frac{2e^2}{h}\frac{1}{3} \sum_\xi\sum_i^3\sum_n^N T_{\xi,i,n},
\end{equation}
where $N=\text{Int}(k_F W/\pi - \tfrac{1}{2})$ is the number of energetically allowed modes set by the contacts' Fermi energy, and the factor $\frac{1}{3}$ accounts for the lifting of the three-fold $K$ and $K'$ point degeneracy in strained graphene. Equations 2, 3, and 4 are similar to previously derived ones\cite{Fogler08,Pellegrino11,Cao12}, but now permit the insertion of experimentally relevant $\Delta \mu_{\text{contact}}$, $\Delta\mu_{\varepsilon}$, $\theta$, and $\bm{A}_{\text{lat},i}$.

We now describe the qualitative impact of uniaxial strain on the transmission, and how it sets the stage for GQSTs. Figure 2(f), $\varepsilon_{\text{total}}= 0$, and (g), $\varepsilon_{\text{total}}\neq 0$, show the Fermi circles in the contacts (big circles) and the channel (small circles). The dashed lines indicate the $y$-momentum conservation imposed to the conduction modes as they propagate from the contacts into the channel. The overlaid solid curves in Fig.\ 2(f) -- (g) show the transmission (radial axis ranges from 0 to 1) as a function of incidence angle on the channel (polar axis). The transmission is calculated using Eq.\ 3 and parameters: $L=100$~nm, $W=1000$~nm, $\Delta\mu_{\text{contact}}=-0.12$ eV, and $\tilde{k}_F=k_F/2.5$. In Fig.\ 2(f), a vector potential of $A_{i,y}=k_F$ is applied. This $A_{i,y}$ splits vertically the channel's Fermi circle into two (one per valley). Concretely, strain modifies the $\phi$ of each transmission mode, and the number of energetically allowed modes. As strain increases, the overlap between the contact (black) and channel (red) Fermi circles shrinks, such that fewer modes permit $y$-momentum conservation. When the circles no longer overlap ($A_{i,y}>k_F+\tilde{k}_F$), transmission is always energetically forbidden and $\sigma \rightarrow 0$. We emphasize that the drop in conductivity does not arise from a strain-induced band gap, which is only expected to occur at much larger uniaxial strains $\sim 20$~\% \cite{Pereira09}.

\section{Graphene Quantum Strain Transistors}
\begin{figure}
\includegraphics{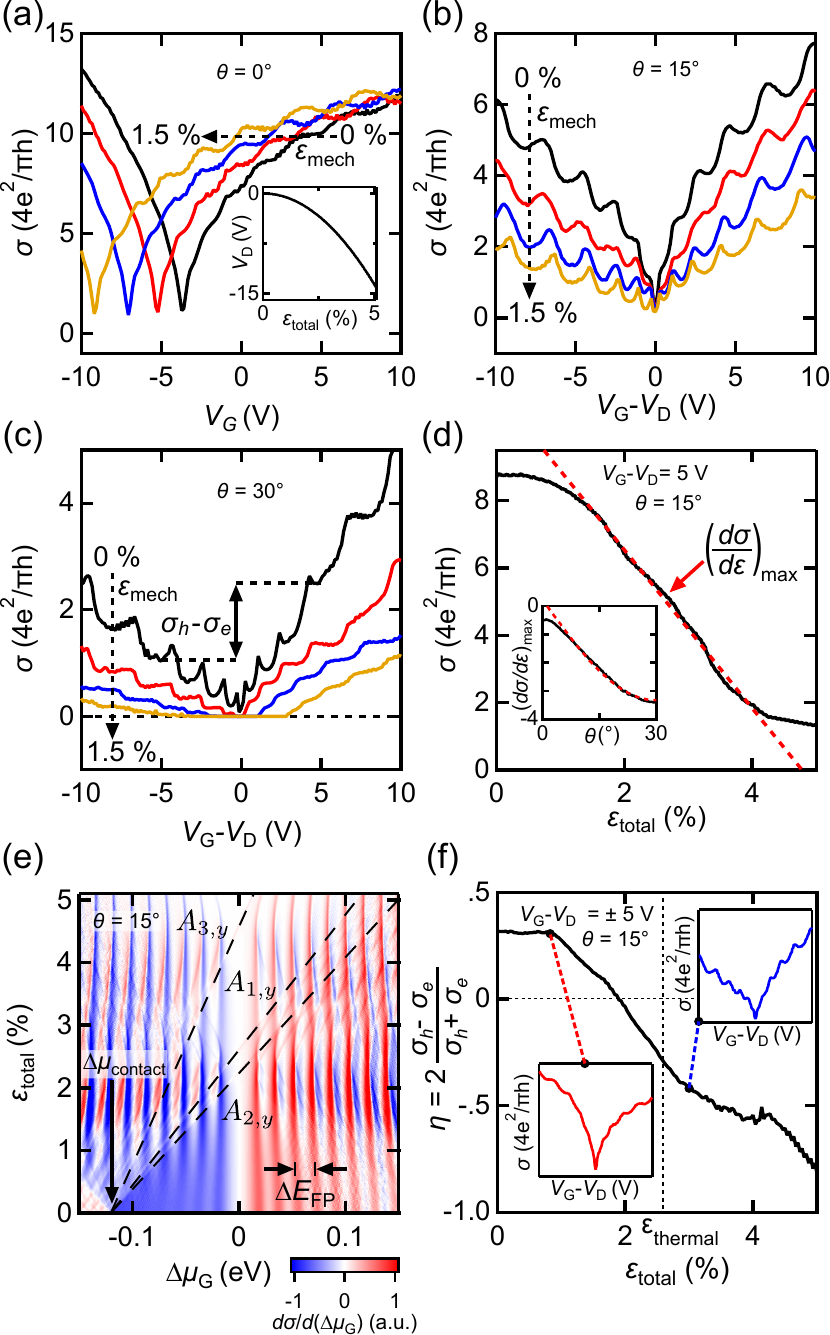}
\caption{\label{Fig3} Conductivity signatures of uniaxial QTSE in graphene. (a) $\sigma$ - $V_{\text{G}}$ for $\varepsilon_{\text{mech}}=$ 0, 0.5, 1.0, and 1.5~$\%$ (black, red, blue, gold) at $\theta=0^{\circ}$. The strain-induced scalar potential shifts the $\sigma$ - $V_{\text{G}}$ curves. The inset shows the gate-shift of the Dirac point, $V_{\text{D}}$, as a function of $\varepsilon_{\text{total}}$. (b) $\sigma$ - $(V_{\text{G}} - V_{\text{D}})$ at $\theta=15 ^{\circ}$. There is a rapid decrease of $\sigma$ with strain. Fabry-P\'erot (FP) resonances are clearly visible. (c) $\sigma$ - $(V_{\text{G}}-V_{\text{D}})$ data at $\theta=30^{\circ}$ show a complete suppression of $\sigma$ under strain. There is an asymmetry between hole ($\sigma_h$) and electron ($\sigma_e$) conductivities at opposite $(V_{\text{G}} - V_{\text{D}})$. (d) $\sigma$ - $\varepsilon_{\text{total}}$ for $\theta=15^{\circ}$, with a linear fit (red) to extract $(d\sigma/d\varepsilon)_{\text{max}}$.  Inset: $(d\sigma/d\varepsilon)_{\text{max}}$ - $\theta$ with a sinusoidal fit (red). (e) Color map of $d\sigma/d(\Delta \mu_{\text{G}})$ versus $\Delta \mu_{\text{G}}$ and $\varepsilon_{\text{total}}$. Clear vertical FP resonances are visible. The dashed lines identify the three conductivity features $k_{F} = \tilde{k}_{F} + A_{i,y}$, where the vector potentials turn off the propagation of conduction modes. (f) Relative electron-hole asymmetry, $\eta=2(\sigma_{h}-\sigma_{e})/(\sigma_{h}+\sigma_{e})$, as a function of $\varepsilon_{\text{total}}$ at $\theta=15 ^{\circ}$.  Insets: $\sigma$ - $ V_{\text{G}}$ shows $\eta>0$ at $\varepsilon_{\text{total}}=1$~\% (bottom left), and $\eta<0$ at $\varepsilon_{\text{total}}=3$~\%  (top right).}
\end{figure}
We set out to map the behavior of the ballistic conductivity from Eq.\ 4 as a function of experimentally tunable parameters. We predict four clear strain-tunable experimental transport signatures as shown in Fig.\ 3: a lateral shift in the $\sigma$ - $V_{\text{G}}$ curves, a dramatic reduction in conductivity, a rich set of ballistic conductivity resonances, and a sizeable electron-hole transport asymmetry. We quantify how these predictions depend on realistic values of the contact doping, crystal orientation, uniaxial strain, and gate voltage.

To calculate the $\sigma$ data in Fig.\ 3 we used the parameters introduced above: $\varepsilon_{\text{thermal}}=2.6$~$\%$, $\Delta\mu_{\text{contact}}=-0.12$~eV, $L=100$~nm and $W=1000$~nm. We considered the experimentally relevant regime where $k_B T\sim 0.1$~meV~$< eV_{\text{B}} < 1$~meV~$\ll |\Delta\mu_{\text{contact}}|= 0.12 $~eV, and $\hbar v_F A_{i,y}$ reaches up to $0.34$ eV at maximum $\varepsilon_{\text{total}} = 5.1$~$\%$, $\theta = 30^{\circ}$. Based on these energy scales, we can safely neglect the minor impact of a small $V_{\text{B}}$ and low temperature on the calculated $\sigma$. We remind the reader that the main objective of our applied theory is to include experimental considerations which have major impacts on the theoretical predictions. For instance, the inclusion of thermally-induced strain $\varepsilon_{\text{thermal}}$ in the channel, the lattice distortion vector potentials $\bm{A}_{\text{lat},i}$, crystal orientation $\theta$, and a realistic contact doping $\Delta \mu_{\text{contact}}$, all lead to order of magnitude changes in $\sigma$ and the transistor on/off ratios. On the other hand, we verified numerically (Supplementary Information S3) that an exhaustive list of other factors such as realistic impurity density, series resistance, gating of $\Delta\mu_{\text{contact}}$, thermal strain in the contacts, uncertainties on $v_{F}$, $\nu$, $\beta$, $L$, $W$, \textit{etc.}, lead to modest or negligible corrections to $\sigma$ in strained devices.

In Fig.\ 3(a), we plot $\sigma$ - $V_{\text{G}}$ for $\varepsilon_{\text{mech}}=$ 0, 0.5, 1.0, and 1.5~$\%$ (black, blue, red, gold) for $\theta = 0^{\circ}$. At this crystal orientation, the $A_{i,y}$ are nearly zero. The only significant consequence of strain is the scalar potential $\Delta\mu_{\varepsilon}=g_{\varepsilon}(1-\nu)\varepsilon_{\text{total}}$. We remind the reader that even when $\varepsilon_{\text{mech}}= 0$~\%, there is a built-in thermal uniaxial strain of $\varepsilon_{\text{thermal}}=2.6$~$\%$. A clear signature of the scalar potential is apparent in Fig.\ 3(a): the $\sigma$ - $V_{\text{G}}$ curves shift with increasing strain. The shift of the gate-position of the Dirac point, $V_{\text{D}}$, is plotted in the inset of Fig.\ 3(a), and given by
$V_{\text{D}}=-\frac{e}{C_{\text{G}}}\frac{g_{\varepsilon}^2}{\pi(\hbar v_F)^2}(1-\nu)^2 \varepsilon_{\text{total}}^2$. The value of $V_{\text{D}}$ is independent of $\theta$ and $\Delta \mu_{\text{contact}}$. It can therefore be used in experiments to measure the $\varepsilon_{\text{mech}}$ generated by the instrumentation, and the built-in $\varepsilon_{\text{thermal}}$.

In Fig.\ 3(b) -- (c), we plot $\sigma$ - $(V_{\text{G}}-V_{\text{D}})$ at various $\varepsilon_{\text{mech}}$. We subtracted $V_{\text{D}}$ from the horizontal axis to remove the lateral shifts arising from the scalar potential, and focus on the effects of the vector potentials. We set $\theta$ respectively to 15$^{\circ}$ and 30$^{\circ}$ in Fig.\ 3(b) and (c). The uniaxial strain rapidly decreases the conductivity, and this suppression is maximized at $\theta = 30 ^{\circ}$. We see that $\sigma$ can reach $\approx$ 0 for $\varepsilon_{\text{mech}}$ as low as 0.5~$\%$. The range of $V_{\text{G}}$ where a clear turning-off of the conductivity is possible grows rapidly as $\varepsilon_{\text{mech}}$ increases to 1.5~$\%$. We remark that the strong tunability of $\sigma$ via uniaxial strain implies that shot noise measurements of the Fano factor would provide yet another signature of the model (Supplementary Information S4).

In Fig.\ 3(d) we show the dependence of $\sigma$ versus $\varepsilon_{\text{total}}$, where $\sigma$ is calculated at ($V_{\text{G}}-V_{\text{D}}) = 5$ V and $\theta = 15 ^{\circ}$. The dashed line is a linear fit of the steepest section of the curve, and defines $(d\sigma/ d\varepsilon)_{\text{max}}$. This latter quantity is plotted and fitted in the inset as a function of $\theta$. The fit is a sinusoidal function matching the form of the hopping vector potential in Eq.\ 2. The discrepancy between the data and fit below $\theta = 5^{\circ}$ indicates that $\bm{A}_{\text{hop}}$ is no longer the main contribution to the $\bm{A}_{i}$ at small angles. The strong, and single-valued, dependence of $(d\sigma/ d\varepsilon)_{\text{max}}$ on $\theta$ makes it plausible to extract the crystal angle from transport data. This relaxes the requirement to measure $\theta$ with polarized Raman spectroscopy\cite{Huang09} or STM imaging \cite{Andrei12}.

Another consequence of uniaxial strain on transport is the shifting of conductivity resonances, which are visible in Fig.\ 3(b), (c) and (e). These resonances arise from interferences of the ballistic carriers as they are transmitted or reflected at the channel-contact interfaces. The Fermi energy spacing of these FP resonances\cite{Rickhaus13} is $\Delta E_{\text{FP}}=\pi \hbar v_{F}/(L\cos\phi$). Figure 3(e) presents $d\sigma/d(\Delta \mu_{\text{G}})$ versus $\Delta \mu_{\text{G}}$ and $\varepsilon_{\text{total}}$ at $\theta = 15^{\circ}$, and shows bright vertical FP resonances. Convoluted with the vertical resonances, are hyperbolas whose asymptotes are labelled with dashed black lines. The slopes of these three clear transport features are related to the individual $A_{i,y}$. Each line is determined by the formula $k_F = \Delta\mu_{\text{contact}}/(\hbar v_{F}) = \Delta\mu_{\text{G}}/(\hbar v_{F}) + A_{i,y}$. The drop in conductivity across these dashed lines corresponds to the closing of conduction modes by the mechanism illustrated in Fig.\ 2(g). The three dashed lines intersect when $\Delta\mu_{\text{contact}} =\Delta\mu_{\text{G}}$. While only $\varepsilon_{\text{total}} > 2.6$~$\%$ would be available in the proposed experiment, the observation and extrapolation of the sharp $A_{i,y}$ features would give a direct measure of both $\Delta\mu_{\text{contact}}$ and $A_{i,y}$.

A fourth effect of $\varepsilon_{\text{mech}}$ is to modify the electron-hole transport asymmetry, $\sigma_{e}\neq\sigma_{h}$. We extract $\sigma_{e}$ and $\sigma_{h}$ at ($V_{\text{G}}-V_{\text{D}}) = \pm$5 V as shown in Fig.\ 3(c). We then define a relative electron-hole transport asymmetry as $\eta=2(\sigma_{h}-\sigma_{e})/(\sigma_{h}+\sigma_{e})$. This relative asymmetry $\eta$ is widely dependent on both $\varepsilon_{\text{mech}}$ and $\theta$. In Fig.\ 3(f) we plot $\eta$ - $\varepsilon_{\text{mech}}$ at $\theta = 15 ^{\circ}$. The two insets display the asymmetric shape of $\sigma$ - $(V_{\text{G}} - V_{\text{D}})$ at respectively $\varepsilon_{\text{total}} =$ 1~$\%$ (bottom left), and 3~$\%$ (top right). At $\varepsilon_{\text{total}}\approx 1.8$~$\%$, the electron-hole asymmetry $\eta$ reverses sign. The vertical dashed line in Fig.\ 3(f) shows the expected $\varepsilon_{\text{thermal}}$ in the devices, and the experimentally available $\varepsilon_{\text{total}}$ values would lie to the right of this line. For a known $\theta$, the dependence of $\eta$ versus $\varepsilon_{\text{total}}$ could be another method to experimentally determine the value of $\Delta\mu_{\text{contact}}$. We are now ready to assess the potential of the proposed devices for a practical demonstration of high on/off ratio graphene quantum strain transistors (GQSTs).

\begin{figure}
\includegraphics{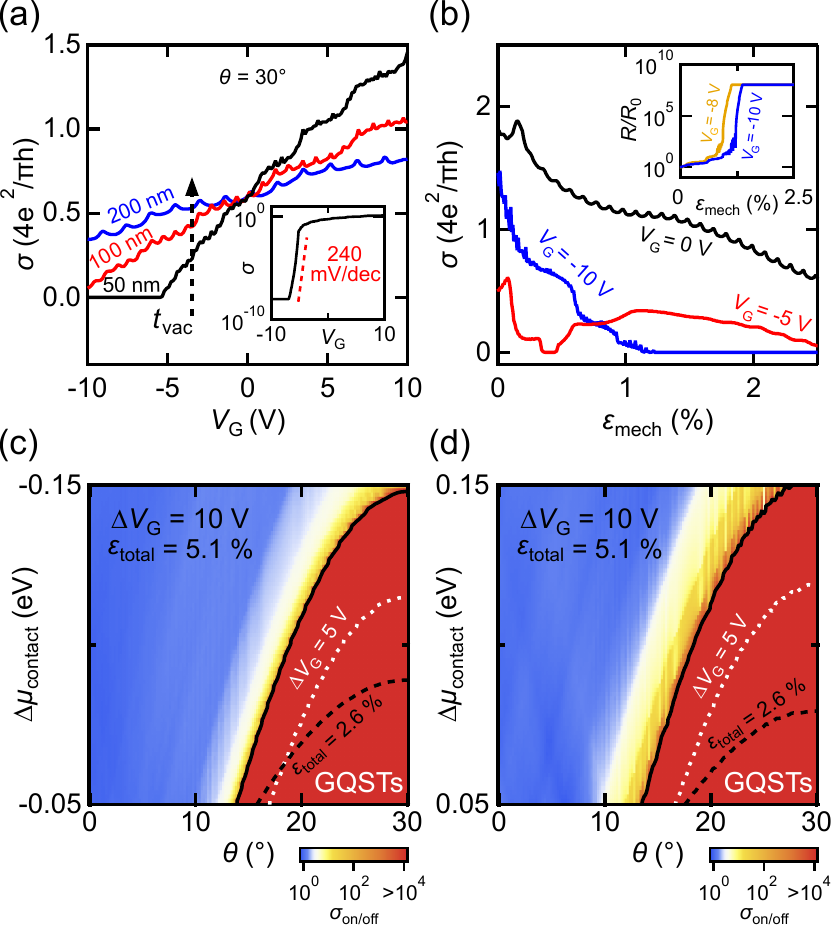}
\caption{Graphene quantum strain transistors (GQSTs). (a) $\sigma$ - $V_{\text{G}}$ with a gate-channel spacing $t_{\text{vac}}=$ 50 nm (black), 100 nm (red), and 200 nm (blue) and under strain $\varepsilon_{\text{total}}=5.1$~$\%$, $\theta = 30^{\circ}$. The 50 nm gate spacing gives rise to a robust GQST effect with $\sigma_{\text{on/off}}\gg10^4$. The inset shows a subthreshold slope of 240 mV/dec when $t_{\text{vac}}=$ 50 nm. (b) $\sigma$ - $\varepsilon_{\text{mech}}$ shows the extreme strain sensitivity of $\sigma$ versus uniaxial strain at fixed $V_{\text{G}} =$ 0 V (black), $-5$ V (red), and $-10$ V (blue).  Inset: Resistance-based strain sensitivity at $V_{\text{G}}= -8$ V (gold) and $-10$ V (blue). (c) $\sigma_{\text{on/off}}$ - $\Delta\mu_{\text{contact}}$ - $\theta$ color maps at $\varepsilon_{\text{total}}=5.1 \%$ when $\Delta\mu_{\text{contact}}<0$ ($p$-doping of contacts),  and (d) when $\Delta\mu_{\text{contact}}>0$ ($n$-doping of contacts). The color scales show $\sigma_{\text{on/off}}$ calculated with $\Delta V_{\text{G}}=$ 10 V (\textit{i.e.} at 0 V and $-$10 V). The contour lines mark the boundary where $\sigma_{\text{on/off}} >10^{4}$, corresponding to a strong GQST effect. The solid black, dashed white, and dashed black contours are for $\Delta V_{\text{G}}=$ 10 V, 5 V, and 10 V but for a lower $\varepsilon_{\text{total}}=2.6$~$\%$, respectively.}
\end{figure}

We consider GQST device operation at low temperature ($\sim$1 K), which imposes a minimum $\varepsilon_{\text{total}} = \varepsilon_{\text{thermal}} =$ 2.6~$\%$ to the suspended channel. Figure 4(a) shows $\sigma$ - $V_{\text{G}}$ at $\varepsilon_{\text{mech}} =$ 2.5~$\%$ for $\theta = 30 ^{\circ}$, and $t_{\text{vac}}$ respectively 50 nm (black), 100 nm (blue), and 200 nm (red). As the spacing is increased, larger $V_{\text{G}}$'s are needed to achieve high on/off ratios. For a realistic $t_{\text{vac}}=$ 50 nm, very high on/off ratios ($\sigma_{\text{on/off}}>10^4$) are possible even for modest $V_{\text{G}}$'s between $ - $5 V and 0 V. The inset of Fig.\ 4(a) plots the $\sigma$ - $V_{\text{G}}$ data on a log scale, and shows a subthreshold slope of 240 mV/dec. We cap the maximum measurable device resistance to $R=100$~G$\Omega$ with $V_{\text{B}}=1$~mV, which corresponds to a $\sigma_{\text{on/off}}= 10^{-8}$ or $R/R_0=10^8$. The subthreshold slope of GQSTs is not limited by the thermionic limit, and would decrease to less than 60 mV/dec when $t_{\text{vac}} <$ 10 nm. In the long term, GQST devices could be designed for room-temperature operation using uniaxial strains built into substrates \cite{Kim09, Zhang18}. This would lift the constraint of suspending the channel.

Figure 4(b) demonstrates a second application for GQSTs, as ultra-sensitive strain or pressure sensors. The turning on and off of the conductivity can be triggered entirely by mechanical actuation, while holding $V_{\text{G}}$ constant. The data are respectively $\sigma$ - $\varepsilon_{\text{mech}}$ at $V_{\text{G}}= 0$~V (black), $ - $5 V (red), and $ - $10 V (blue), with $t_{\text{vac}} =$ 50 nm and $\theta = 30 ^{\circ}$. In the inset of Fig.\ 4(b), we show the expected relative resistance change, $R/R_0$, of the graphene channel at $V_{\text{G}} = -$8 V and $-$10 V. While previously demonstrated graphene strain sensors\cite{Smith16,Chun17} have shown a roughly linear resistance change with strain, the present devices show a range of exponential sensitivity. The location in $\varepsilon_{\text{mech}}$ of this exponential sensitivity region is tunable with $V_{\text{G}}$.

In Figs.\ 4(c) -- (d), we plot color maps of the highest $\sigma_{\text{on/off}}$ ratio extracted from data similar to Fig.\ 4(a) within a range $\Delta V_{\text{G}} = 10$ V (\textit{i.e.} comparing $\sigma$ at $V_{\text{G}} = 0$ V and $-10$ V). We map the on/off ratio, for $t_{\text{vac}} =$ 50 nm, as a function of the lattice orientation $\theta$, and the contact doping $\Delta\mu_{\text{contact}} < 0$ and $> 0$, in panel (c) and (d) respectively. There is a broad region (red) of parameter space predicting excellent GQSTs, making it plausible that this platform can lead to an experimental demonstration. For instance, when $\varepsilon_{\text{total}}=$ 5.1~$\%$ and $|\Delta\mu_{\text{contact}}| \sim 0.1$ eV, a $\sigma_{\text{on/off}}>10^4$ is achieved in devices where $\theta \geq 18^{\circ}$. Moreover, an excellent GQST effect is also calculated over a broad parameter range when using a smaller $\Delta V_{\text{G}} = 5$ V ($V_{\text{G}} = 0$ V and $-5$ V), as shown by the dashed white contours. Finally, the effect is robust to variations in the exact value of the applied mechanical and thermal strains. To show this, we draw additional dashed black contour lines in Figs.\ 4(c) -- (d) corresponding to $\sigma_{\text{on/off}}>10^4$ when $\varepsilon_{\text{total}}=$ 2.6~$\%$.

We note that a series resistance arising from the gold-graphene interface, or an extrinsic impurity doping of the channel would not significantly modify Fig.\ 4 (c) -- (d). For a realistic series resistance\cite{Sundaram11,Anzi18} of 400 $\Omega$, the calculated $\sigma$ at $\varepsilon_{\text{total}}=$ 5.1~$\%$ is modified by less than 20~$\%$ (Supplementary Information S3), and has a negligible impact on $\sigma_{\text{on/off}}$. A realistic charge impurity density \cite{Yigen14} of $n_{\text{imp}} = 5\times 10^{10}$ cm$^{-2}$ is equivalent to the doping created by ($V_{\text{G}}-V_{\text{D}}) =$ 0.45 V, for $t_{\text{vac}}=$ 50 nm. From the inset of Fig.\ 3(a), we see that the location of the charge degeneracy point $V_{\text{D}}$ is significantly below $V_{\text{G}}= -10$ V at $\varepsilon_{\text{total}} = 5.1$~$\%$. Consequently, in Fig.\ 4 the doping of the channel is largely dominated by the electrostatic gate doping, and the impact of $n_{\text{imp}}$ would be negligible.

\section{Conclusions}

In conclusion, we proposed an experimental platform able to realistically implement proposals for uniaxial strain-engineering of quantum transport in 2D materials. Some of the key aspects of the proposed devices and instrumentation are a wide aspect ratio of the transistor channels, removing the need for ordered crystal edges, and independent control of both the mechanical strain and the charge density in the devices. This platform provides a wide experimental tunability for low-temperature transport experiments.

We then reported an applied theoretical model which quantitatively describes uniaxial QTSE in graphene. The model includes the effect of strain on the Fermi energy (scalar potential), on the position of the Dirac points in $k$-space (gauge vector potentials), and the anisotropy of the Fermi velocity. We included the dominant experimental parameters such as, contact doping, the reciprocal lattice distortion for arbitrary crystal orientations, and both the thermally and mechanically generated strains. We showed that the $\sigma$ - $V_{\text{G}}$ calculated data as a function of $\varepsilon_{\text{total}}$ give four experimentally observable signatures. They are a gate-shift of the charge degeneracy point, a dramatic decrease in conductivity, a rich set of ballistic interferences, and a tunable electron-hole conductivity asymmetry.

Finally, we assessed the performance of the proposed GQST devices. We found that comfortably within the experimental capabilities, the charge conductivity in graphene devices can be completely suppressed by uniaxial strain. We mapped out the parameter space for contact doping, crystal orientation, and applied strain where a robust GQST effect with $\sigma_{\text{on/off}}>10^4$ can be achieved using modest $V_{\text{G}}$'s. This transistor effect is purely a result of the quantum (ballistic) nature of the transport, and does not arise from band gap generation. Recent progress in making room-temperature ballistic graphene devices \cite{Banzerus16}, and building controlled strain fields into substrates \cite{Kim09, Zhang18}, pave the way for GQSTs to be used in flexible electronics\cite{Akinwande14}, and in valleytronics \cite{Fujita10,Yesilyurt16}. More immediately, the proposed platform and model should lead to a demonstration and optimization of GQSTs.

This work was supported by NSERC (Canada), CFI (Canada), and Concordia University. We acknowledge using the LMF cleanroom (Laboratoire de Microfabrication) at Polytechnique Montr\'{e}al.


\begin{thebibliography}{49}%
\makeatletter
\providecommand \@ifxundefined [1]{%
 \@ifx{#1\undefined}
}%
\providecommand \@ifnum [1]{%
 \ifnum #1\expandafter \@firstoftwo
 \else \expandafter \@secondoftwo
 \fi
}%
\providecommand \@ifx [1]{%
 \ifx #1\expandafter \@firstoftwo
 \else \expandafter \@secondoftwo
 \fi
}%
\providecommand \natexlab [1]{#1}%
\providecommand \enquote  [1]{``#1''}%
\providecommand \bibnamefont  [1]{#1}%
\providecommand \bibfnamefont [1]{#1}%
\providecommand \citenamefont [1]{#1}%
\providecommand \href@noop [0]{\@secondoftwo}%
\providecommand \href [0]{\begingroup \@sanitize@url \@href}%
\providecommand \@href[1]{\@@startlink{#1}\@@href}%
\providecommand \@@href[1]{\endgroup#1\@@endlink}%
\providecommand \@sanitize@url [0]{\catcode `\\12\catcode `\$12\catcode
  `\&12\catcode `\#12\catcode `\^12\catcode `\_12\catcode `\%12\relax}%
\providecommand \@@startlink[1]{}%
\providecommand \@@endlink[0]{}%
\providecommand \url  [0]{\begingroup\@sanitize@url \@url }%
\providecommand \@url [1]{\endgroup\@href {#1}{\urlprefix }}%
\providecommand \urlprefix  [0]{URL }%
\providecommand \Eprint [0]{\href }%
\providecommand \doibase [0]{http://dx.doi.org/}%
\providecommand \selectlanguage [0]{\@gobble}%
\providecommand \bibinfo  [0]{\@secondoftwo}%
\providecommand \bibfield  [0]{\@secondoftwo}%
\providecommand \translation [1]{[#1]}%
\providecommand \BibitemOpen [0]{}%
\providecommand \bibitemStop [0]{}%
\providecommand \bibitemNoStop [0]{.\EOS\space}%
\providecommand \EOS [0]{\spacefactor3000\relax}%
\providecommand \BibitemShut  [1]{\csname bibitem#1\endcsname}%
\let\auto@bib@innerbib\@empty
\bibitem [{\citenamefont {Fogler}\ \emph {et~al.}(2008)\citenamefont {Fogler},
  \citenamefont {Guinea},\ and\ \citenamefont {Katsnelson}}]{Fogler08}%
  \BibitemOpen
  \bibfield  {author} {\bibinfo {author} {\bibfnamefont {M.~M.}\ \bibnamefont
  {Fogler}}, \bibinfo {author} {\bibfnamefont {F.}~\bibnamefont {Guinea}}, \
  and\ \bibinfo {author} {\bibfnamefont {M.~I.}\ \bibnamefont {Katsnelson}},\
  }\href@noop {} {\bibfield  {journal} {\bibinfo  {journal} {Phys. Rev. Lett.}\
  }\textbf {\bibinfo {volume} {101}},\ \bibinfo {pages} {226804} (\bibinfo
  {year} {2008})}\BibitemShut {NoStop}%
\bibitem [{\citenamefont {Pellegrino}\ \emph {et~al.}(2011)\citenamefont
  {Pellegrino}, \citenamefont {Angilella},\ and\ \citenamefont
  {Pucci}}]{Pellegrino11}%
  \BibitemOpen
  \bibfield  {author} {\bibinfo {author} {\bibfnamefont {F.~M.~D.}\
  \bibnamefont {Pellegrino}}, \bibinfo {author} {\bibfnamefont {G.~G.~N.}\
  \bibnamefont {Angilella}}, \ and\ \bibinfo {author} {\bibfnamefont
  {R.}~\bibnamefont {Pucci}},\ }\href@noop {} {\bibfield  {journal} {\bibinfo
  {journal} {Phys. Rev. B}\ }\textbf {\bibinfo {volume} {84}},\ \bibinfo
  {pages} {195404} (\bibinfo {year} {2011})}\BibitemShut {NoStop}%
\bibitem [{\citenamefont {Cao}\ \emph {et~al.}(2012)\citenamefont {Cao},
  \citenamefont {Cheng},\ and\ \citenamefont {Li}}]{Cao12}%
  \BibitemOpen
  \bibfield  {author} {\bibinfo {author} {\bibfnamefont {Z.-Z.}\ \bibnamefont
  {Cao}}, \bibinfo {author} {\bibfnamefont {Y.-F.}\ \bibnamefont {Cheng}}, \
  and\ \bibinfo {author} {\bibfnamefont {G.-Q.}\ \bibnamefont {Li}},\
  }\href@noop {} {\bibfield  {journal} {\bibinfo  {journal} {Appl. Phys.
  Lett.}\ }\textbf {\bibinfo {volume} {101}},\ \bibinfo {pages} {253507}
  (\bibinfo {year} {2012})}\BibitemShut {NoStop}%
\bibitem [{\citenamefont {Castro~Neto}\ \emph {et~al.}(2009)\citenamefont
  {Castro~Neto}, \citenamefont {Guinea}, \citenamefont {Peres}, \citenamefont
  {Novoselov},\ and\ \citenamefont {Geim}}]{CastroNeto09}%
  \BibitemOpen
  \bibfield  {author} {\bibinfo {author} {\bibfnamefont {A.~H.}\ \bibnamefont
  {Castro~Neto}}, \bibinfo {author} {\bibfnamefont {F.}~\bibnamefont {Guinea}},
  \bibinfo {author} {\bibfnamefont {N.~M.~R.}\ \bibnamefont {Peres}}, \bibinfo
  {author} {\bibfnamefont {K.~S.}\ \bibnamefont {Novoselov}}, \ and\ \bibinfo
  {author} {\bibfnamefont {A.~K.}\ \bibnamefont {Geim}},\ }\href@noop {}
  {\bibfield  {journal} {\bibinfo  {journal} {Rev. Mod. Phys.}\ }\textbf
  {\bibinfo {volume} {81}},\ \bibinfo {pages} {109} (\bibinfo {year}
  {2009})}\BibitemShut {NoStop}%
\bibitem [{\citenamefont {Mayorov}\ \emph {et~al.}(2011)\citenamefont
  {Mayorov}, \citenamefont {Gorbachev}, \citenamefont {Morozov}, \citenamefont
  {Britnell}, \citenamefont {Jalil}, \citenamefont {Ponomarenko}, \citenamefont
  {Blake}, \citenamefont {Novoselov}, \citenamefont {Watanabe}, \citenamefont
  {Taniguchi},\ and\ \citenamefont {Geim}}]{Mayorov11}%
  \BibitemOpen
  \bibfield  {author} {\bibinfo {author} {\bibfnamefont {A.~S.}\ \bibnamefont
  {Mayorov}}, \bibinfo {author} {\bibfnamefont {R.~V.}\ \bibnamefont
  {Gorbachev}}, \bibinfo {author} {\bibfnamefont {S.~V.}\ \bibnamefont
  {Morozov}}, \bibinfo {author} {\bibfnamefont {L.}~\bibnamefont {Britnell}},
  \bibinfo {author} {\bibfnamefont {R.}~\bibnamefont {Jalil}}, \bibinfo
  {author} {\bibfnamefont {L.~A.}\ \bibnamefont {Ponomarenko}}, \bibinfo
  {author} {\bibfnamefont {P.}~\bibnamefont {Blake}}, \bibinfo {author}
  {\bibfnamefont {K.~S.}\ \bibnamefont {Novoselov}}, \bibinfo {author}
  {\bibfnamefont {K.}~\bibnamefont {Watanabe}}, \bibinfo {author}
  {\bibfnamefont {T.}~\bibnamefont {Taniguchi}}, \ and\ \bibinfo {author}
  {\bibfnamefont {A.~K.}\ \bibnamefont {Geim}},\ }\href@noop {} {\bibfield
  {journal} {\bibinfo  {journal} {Nano Lett.}\ }\textbf {\bibinfo {volume}
  {11}},\ \bibinfo {pages} {2396} (\bibinfo {year} {2011})}\BibitemShut
  {NoStop}%
\bibitem [{\citenamefont {Banszerus}\ \emph {et~al.}(2016)\citenamefont
  {Banszerus}, \citenamefont {Schmitz}, \citenamefont {Engels}, \citenamefont
  {Goldsche}, \citenamefont {Watanabe}, \citenamefont {Taniguchi},
  \citenamefont {Beschoten},\ and\ \citenamefont {Stampfer}}]{Banzerus16}%
  \BibitemOpen
  \bibfield  {author} {\bibinfo {author} {\bibfnamefont {L.}~\bibnamefont
  {Banszerus}}, \bibinfo {author} {\bibfnamefont {M.}~\bibnamefont {Schmitz}},
  \bibinfo {author} {\bibfnamefont {S.}~\bibnamefont {Engels}}, \bibinfo
  {author} {\bibfnamefont {M.}~\bibnamefont {Goldsche}}, \bibinfo {author}
  {\bibfnamefont {K.}~\bibnamefont {Watanabe}}, \bibinfo {author}
  {\bibfnamefont {T.}~\bibnamefont {Taniguchi}}, \bibinfo {author}
  {\bibfnamefont {B.}~\bibnamefont {Beschoten}}, \ and\ \bibinfo {author}
  {\bibfnamefont {C.}~\bibnamefont {Stampfer}},\ }\href@noop {} {\bibfield
  {journal} {\bibinfo  {journal} {Nano Lett.}\ }\textbf {\bibinfo {volume}
  {16}},\ \bibinfo {pages} {1387} (\bibinfo {year} {2016})}\BibitemShut
  {NoStop}%
\bibitem [{\citenamefont {Akinwande}\ \emph {et~al.}(2014)\citenamefont
  {Akinwande}, \citenamefont {Petrone},\ and\ \citenamefont
  {Hone}}]{Akinwande14}%
  \BibitemOpen
  \bibfield  {author} {\bibinfo {author} {\bibfnamefont {D.}~\bibnamefont
  {Akinwande}}, \bibinfo {author} {\bibfnamefont {N.}~\bibnamefont {Petrone}},
  \ and\ \bibinfo {author} {\bibfnamefont {J.}~\bibnamefont {Hone}},\
  }\href@noop {} {\bibfield  {journal} {\bibinfo  {journal} {Nat. Commun.}\
  }\textbf {\bibinfo {volume} {5}},\ \bibinfo {pages} {5678} (\bibinfo {year}
  {2014})}\BibitemShut {NoStop}%
\bibitem [{\citenamefont {Iannaccone}\ \emph {et~al.}(2018)\citenamefont
  {Iannaccone}, \citenamefont {Bonaccorso}, \citenamefont {Colombo},\ and\
  \citenamefont {Fiori}}]{Iannaccone18}%
  \BibitemOpen
  \bibfield  {author} {\bibinfo {author} {\bibfnamefont {G.}~\bibnamefont
  {Iannaccone}}, \bibinfo {author} {\bibfnamefont {F.}~\bibnamefont
  {Bonaccorso}}, \bibinfo {author} {\bibfnamefont {L.}~\bibnamefont {Colombo}},
  \ and\ \bibinfo {author} {\bibfnamefont {G.}~\bibnamefont {Fiori}},\
  }\href@noop {} {\bibfield  {journal} {\bibinfo  {journal} {Nat.
  Nanotechnol.}\ }\textbf {\bibinfo {volume} {13}},\ \bibinfo {pages} {183}
  (\bibinfo {year} {2018})}\BibitemShut {NoStop}%
\bibitem [{\citenamefont {Guinea}\ \emph {et~al.}(2010)\citenamefont {Guinea},
  \citenamefont {Katsnelson},\ and\ \citenamefont {Geim}}]{Guinea10}%
  \BibitemOpen
  \bibfield  {author} {\bibinfo {author} {\bibfnamefont {F.}~\bibnamefont
  {Guinea}}, \bibinfo {author} {\bibfnamefont {M.~I.}\ \bibnamefont
  {Katsnelson}}, \ and\ \bibinfo {author} {\bibfnamefont {A.~K.}\ \bibnamefont
  {Geim}},\ }\href@noop {} {\bibfield  {journal} {\bibinfo  {journal} {Nat.
  Phys.}\ }\textbf {\bibinfo {volume} {6}},\ \bibinfo {pages} {30} (\bibinfo
  {year} {2010})}\BibitemShut {NoStop}%
\bibitem [{\citenamefont {Amorim}\ \emph {et~al.}(2016)\citenamefont {Amorim},
  \citenamefont {Cortijo}, \citenamefont {de~Juan}, \citenamefont {Grushine},
  \citenamefont {Guinea}, \citenamefont {Gutierrez-Rubio}, \citenamefont
  {Ochoa}, \citenamefont {Parente}, \citenamefont {Roldan}, \citenamefont
  {San-Jose}, \citenamefont {Schiefele}, \citenamefont {Sturla},\ and\
  \citenamefont {Vozmediano}}]{Amorim16}%
  \BibitemOpen
  \bibfield  {author} {\bibinfo {author} {\bibfnamefont {B.}~\bibnamefont
  {Amorim}}, \bibinfo {author} {\bibfnamefont {A.}~\bibnamefont {Cortijo}},
  \bibinfo {author} {\bibfnamefont {F.}~\bibnamefont {de~Juan}}, \bibinfo
  {author} {\bibfnamefont {A.~G.}\ \bibnamefont {Grushine}}, \bibinfo {author}
  {\bibfnamefont {F.}~\bibnamefont {Guinea}}, \bibinfo {author} {\bibfnamefont
  {A.}~\bibnamefont {Gutierrez-Rubio}}, \bibinfo {author} {\bibfnamefont
  {H.}~\bibnamefont {Ochoa}}, \bibinfo {author} {\bibfnamefont
  {V.}~\bibnamefont {Parente}}, \bibinfo {author} {\bibfnamefont
  {R.}~\bibnamefont {Roldan}}, \bibinfo {author} {\bibfnamefont
  {P.}~\bibnamefont {San-Jose}}, \bibinfo {author} {\bibfnamefont
  {J.}~\bibnamefont {Schiefele}}, \bibinfo {author} {\bibfnamefont
  {M.}~\bibnamefont {Sturla}}, \ and\ \bibinfo {author} {\bibfnamefont
  {M.~A.~H.}\ \bibnamefont {Vozmediano}},\ }\href@noop {} {\bibfield  {journal}
  {\bibinfo  {journal} {Phys. Rep.}\ }\textbf {\bibinfo {volume} {617}},\
  \bibinfo {pages} {1} (\bibinfo {year} {2016})}\BibitemShut {NoStop}%
\bibitem [{\citenamefont {Settnes}\ \emph {et~al.}(2017)\citenamefont
  {Settnes}, \citenamefont {Garcia},\ and\ \citenamefont {Roche}}]{Settnes17}%
  \BibitemOpen
  \bibfield  {author} {\bibinfo {author} {\bibfnamefont {M.}~\bibnamefont
  {Settnes}}, \bibinfo {author} {\bibfnamefont {J.~H.}\ \bibnamefont {Garcia}},
  \ and\ \bibinfo {author} {\bibfnamefont {S.}~\bibnamefont {Roche}},\
  }\href@noop {} {\bibfield  {journal} {\bibinfo  {journal} {2D Mater.}\
  }\textbf {\bibinfo {volume} {4}},\ \bibinfo {pages} {031006} (\bibinfo {year}
  {2017})}\BibitemShut {NoStop}%
\bibitem [{\citenamefont {Wu}\ \emph {et~al.}(2017)\citenamefont {Wu},
  \citenamefont {Shi}, \citenamefont {Sreejith},\ and\ \citenamefont
  {Liu}}]{Wu17}%
  \BibitemOpen
  \bibfield  {author} {\bibinfo {author} {\bibfnamefont {Y.-H.}\ \bibnamefont
  {Wu}}, \bibinfo {author} {\bibfnamefont {T.}~\bibnamefont {Shi}}, \bibinfo
  {author} {\bibfnamefont {G.~J.}\ \bibnamefont {Sreejith}}, \ and\ \bibinfo
  {author} {\bibfnamefont {Z.-X.}\ \bibnamefont {Liu}},\ }\href@noop {}
  {\bibfield  {journal} {\bibinfo  {journal} {Phys. Rev. B}\ }\textbf {\bibinfo
  {volume} {96}},\ \bibinfo {pages} {085138} (\bibinfo {year}
  {2017})}\BibitemShut {NoStop}%
\bibitem [{\citenamefont {Naumis}\ \emph {et~al.}(2017)\citenamefont {Naumis},
  \citenamefont {Barraza-Lopez}, \citenamefont {Oliva-Leyva},\ and\
  \citenamefont {Terrones}}]{Naumis17}%
  \BibitemOpen
  \bibfield  {author} {\bibinfo {author} {\bibfnamefont {G.~G.}\ \bibnamefont
  {Naumis}}, \bibinfo {author} {\bibfnamefont {S.}~\bibnamefont
  {Barraza-Lopez}}, \bibinfo {author} {\bibfnamefont {M.}~\bibnamefont
  {Oliva-Leyva}}, \ and\ \bibinfo {author} {\bibfnamefont {H.}~\bibnamefont
  {Terrones}},\ }\href@noop {} {\bibfield  {journal} {\bibinfo  {journal} {Rep.
  Prog. Phys.}\ }\textbf {\bibinfo {volume} {80}},\ \bibinfo {pages} {096501}
  (\bibinfo {year} {2017})}\BibitemShut {NoStop}%
\bibitem [{\citenamefont {Kitt}\ \emph {et~al.}(2012)\citenamefont {Kitt},
  \citenamefont {Pereira}, \citenamefont {Swan},\ and\ \citenamefont
  {Goldberg}}]{Kitt12}%
  \BibitemOpen
  \bibfield  {author} {\bibinfo {author} {\bibfnamefont {A.~L.}\ \bibnamefont
  {Kitt}}, \bibinfo {author} {\bibfnamefont {V.~M.}\ \bibnamefont {Pereira}},
  \bibinfo {author} {\bibfnamefont {A.~K.}\ \bibnamefont {Swan}}, \ and\
  \bibinfo {author} {\bibfnamefont {B.~B.}\ \bibnamefont {Goldberg}},\
  }\href@noop {} {\bibfield  {journal} {\bibinfo  {journal} {Phys. Rev. B}\
  }\textbf {\bibinfo {volume} {85}},\ \bibinfo {pages} {115432} (\bibinfo
  {year} {2012})}\BibitemShut {NoStop}%
\bibitem [{\citenamefont {Kitt}\ \emph {et~al.}(2013)\citenamefont {Kitt},
  \citenamefont {Pereira}, \citenamefont {Swan},\ and\ \citenamefont
  {Goldberg}}]{Kitt13}%
  \BibitemOpen
  \bibfield  {author} {\bibinfo {author} {\bibfnamefont {A.~L.}\ \bibnamefont
  {Kitt}}, \bibinfo {author} {\bibfnamefont {V.~M.}\ \bibnamefont {Pereira}},
  \bibinfo {author} {\bibfnamefont {A.~K.}\ \bibnamefont {Swan}}, \ and\
  \bibinfo {author} {\bibfnamefont {B.~B.}\ \bibnamefont {Goldberg}},\
  }\href@noop {} {\bibfield  {journal} {\bibinfo  {journal} {Phys. Rev. B}\
  }\textbf {\bibinfo {volume} {87}},\ \bibinfo {pages} {159909} (\bibinfo
  {year} {2013})}\BibitemShut {NoStop}%
\bibitem [{\citenamefont {Levy}\ \emph {et~al.}(2010)\citenamefont {Levy},
  \citenamefont {Burke}, \citenamefont {Meaker}, \citenamefont {Panlasigui},
  \citenamefont {Zettl}, \citenamefont {Guinea}, \citenamefont {Castro~Neto},\
  and\ \citenamefont {Crommie}}]{Levy10}%
  \BibitemOpen
  \bibfield  {author} {\bibinfo {author} {\bibfnamefont {N.}~\bibnamefont
  {Levy}}, \bibinfo {author} {\bibfnamefont {S.~A.}\ \bibnamefont {Burke}},
  \bibinfo {author} {\bibfnamefont {K.~L.}\ \bibnamefont {Meaker}}, \bibinfo
  {author} {\bibfnamefont {M.}~\bibnamefont {Panlasigui}}, \bibinfo {author}
  {\bibfnamefont {A.}~\bibnamefont {Zettl}}, \bibinfo {author} {\bibfnamefont
  {F.}~\bibnamefont {Guinea}}, \bibinfo {author} {\bibfnamefont {A.~H.}\
  \bibnamefont {Castro~Neto}}, \ and\ \bibinfo {author} {\bibfnamefont {M.~F.}\
  \bibnamefont {Crommie}},\ }\href@noop {} {\bibfield  {journal} {\bibinfo
  {journal} {Science}\ }\textbf {\bibinfo {volume} {329}},\ \bibinfo {pages}
  {544} (\bibinfo {year} {2010})}\BibitemShut {NoStop}%
\bibitem [{\citenamefont {Kim}\ \emph {et~al.}(2009)\citenamefont {Kim},
  \citenamefont {Zhao}, \citenamefont {Jang}, \citenamefont {Lee},
  \citenamefont {Kim}, \citenamefont {Kim}, \citenamefont {Ahn}, \citenamefont
  {Kim}, \citenamefont {Choi},\ and\ \citenamefont {Hong}}]{Kim09}%
  \BibitemOpen
  \bibfield  {author} {\bibinfo {author} {\bibfnamefont {K.~S.}\ \bibnamefont
  {Kim}}, \bibinfo {author} {\bibfnamefont {Y.}~\bibnamefont {Zhao}}, \bibinfo
  {author} {\bibfnamefont {H.}~\bibnamefont {Jang}}, \bibinfo {author}
  {\bibfnamefont {S.~Y.}\ \bibnamefont {Lee}}, \bibinfo {author} {\bibfnamefont
  {J.~M.}\ \bibnamefont {Kim}}, \bibinfo {author} {\bibfnamefont {K.~S.}\
  \bibnamefont {Kim}}, \bibinfo {author} {\bibfnamefont {J.~H.}\ \bibnamefont
  {Ahn}}, \bibinfo {author} {\bibfnamefont {P.}~\bibnamefont {Kim}}, \bibinfo
  {author} {\bibfnamefont {J.~Y.}\ \bibnamefont {Choi}}, \ and\ \bibinfo
  {author} {\bibfnamefont {B.~H.}\ \bibnamefont {Hong}},\ }\href@noop {}
  {\bibfield  {journal} {\bibinfo  {journal} {Nature}\ }\textbf {\bibinfo
  {volume} {457}},\ \bibinfo {pages} {706} (\bibinfo {year}
  {2009})}\BibitemShut {NoStop}%
\bibitem [{\citenamefont {Zhang}\ \emph {et~al.}(2018)\citenamefont {Zhang},
  \citenamefont {Heiranian}, \citenamefont {Janicek}, \citenamefont {Budrikis},
  \citenamefont {Zapperi}, \citenamefont {Huang}, \citenamefont {Johnson},
  \citenamefont {Aluru}, \citenamefont {Lyding},\ and\ \citenamefont
  {Mason}}]{Zhang18}%
  \BibitemOpen
  \bibfield  {author} {\bibinfo {author} {\bibfnamefont {Y.}~\bibnamefont
  {Zhang}}, \bibinfo {author} {\bibfnamefont {M.}~\bibnamefont {Heiranian}},
  \bibinfo {author} {\bibfnamefont {B.}~\bibnamefont {Janicek}}, \bibinfo
  {author} {\bibfnamefont {Z.}~\bibnamefont {Budrikis}}, \bibinfo {author}
  {\bibfnamefont {S.}~\bibnamefont {Zapperi}}, \bibinfo {author} {\bibfnamefont
  {P.~Y.}\ \bibnamefont {Huang}}, \bibinfo {author} {\bibfnamefont {H.~T.}\
  \bibnamefont {Johnson}}, \bibinfo {author} {\bibfnamefont {N.~R.}\
  \bibnamefont {Aluru}}, \bibinfo {author} {\bibfnamefont {J.~W.}\ \bibnamefont
  {Lyding}}, \ and\ \bibinfo {author} {\bibfnamefont {N.}~\bibnamefont
  {Mason}},\ }\href@noop {} {\bibfield  {journal} {\bibinfo  {journal} {Nano
  Lett.}\ }\textbf {\bibinfo {volume} {18}},\ \bibinfo {pages} {2098} (\bibinfo
  {year} {2018})}\BibitemShut {NoStop}%
\bibitem [{\citenamefont {Guan}\ and\ \citenamefont {Du}(2017)}]{Guan17}%
  \BibitemOpen
  \bibfield  {author} {\bibinfo {author} {\bibfnamefont {F.}~\bibnamefont
  {Guan}}\ and\ \bibinfo {author} {\bibfnamefont {X.}~\bibnamefont {Du}},\
  }\href@noop {} {\bibfield  {journal} {\bibinfo  {journal} {Nano Lett.}\
  }\textbf {\bibinfo {volume} {17}},\ \bibinfo {pages} {7009} (\bibinfo {year}
  {2017})}\BibitemShut {NoStop}%
\bibitem [{\citenamefont {Tan}\ \emph {et~al.}(2017)\citenamefont {Tan},
  \citenamefont {Elahi}, \citenamefont {Tsao}, \citenamefont {Habib},
  \citenamefont {Barker},\ and\ \citenamefont {Ghosh}}]{Tan17}%
  \BibitemOpen
  \bibfield  {author} {\bibinfo {author} {\bibfnamefont {Y.~H.}\ \bibnamefont
  {Tan}}, \bibinfo {author} {\bibfnamefont {M.~M.}\ \bibnamefont {Elahi}},
  \bibinfo {author} {\bibfnamefont {H.~Y.}\ \bibnamefont {Tsao}}, \bibinfo
  {author} {\bibfnamefont {K.~M.~M.}\ \bibnamefont {Habib}}, \bibinfo {author}
  {\bibfnamefont {N.~S.}\ \bibnamefont {Barker}}, \ and\ \bibinfo {author}
  {\bibfnamefont {A.~W.}\ \bibnamefont {Ghosh}},\ }\href@noop {} {\bibfield
  {journal} {\bibinfo  {journal} {Sci. Rep.}\ }\textbf {\bibinfo {volume}
  {7}},\ \bibinfo {pages} {9714} (\bibinfo {year} {2017})}\BibitemShut
  {NoStop}%
\bibitem [{\citenamefont {Smith}\ \emph {et~al.}(2016)\citenamefont {Smith},
  \citenamefont {Niklaus}, \citenamefont {Paussa}, \citenamefont {Schroder},
  \citenamefont {Fischer}, \citenamefont {Sterner}, \citenamefont {Wagner},
  \citenamefont {Vaziri}, \citenamefont {Forsberg}, \citenamefont {Esseni},
  \citenamefont {Ostling},\ and\ \citenamefont {Lemme}}]{Smith16}%
  \BibitemOpen
  \bibfield  {author} {\bibinfo {author} {\bibfnamefont {A.~D.}\ \bibnamefont
  {Smith}}, \bibinfo {author} {\bibfnamefont {F.}~\bibnamefont {Niklaus}},
  \bibinfo {author} {\bibfnamefont {A.}~\bibnamefont {Paussa}}, \bibinfo
  {author} {\bibfnamefont {S.}~\bibnamefont {Schroder}}, \bibinfo {author}
  {\bibfnamefont {A.~C.}\ \bibnamefont {Fischer}}, \bibinfo {author}
  {\bibfnamefont {M.}~\bibnamefont {Sterner}}, \bibinfo {author} {\bibfnamefont
  {S.}~\bibnamefont {Wagner}}, \bibinfo {author} {\bibfnamefont
  {S.}~\bibnamefont {Vaziri}}, \bibinfo {author} {\bibfnamefont
  {F.}~\bibnamefont {Forsberg}}, \bibinfo {author} {\bibfnamefont
  {D.}~\bibnamefont {Esseni}}, \bibinfo {author} {\bibfnamefont
  {M.}~\bibnamefont {Ostling}}, \ and\ \bibinfo {author} {\bibfnamefont
  {M.~C.}\ \bibnamefont {Lemme}},\ }\href@noop {} {\bibfield  {journal}
  {\bibinfo  {journal} {ACS Nano}\ }\textbf {\bibinfo {volume} {10}},\ \bibinfo
  {pages} {9879} (\bibinfo {year} {2016})}\BibitemShut {NoStop}%
\bibitem [{\citenamefont {Chun}\ \emph {et~al.}(2017)\citenamefont {Chun},
  \citenamefont {Choi},\ and\ \citenamefont {Park}}]{Chun17}%
  \BibitemOpen
  \bibfield  {author} {\bibinfo {author} {\bibfnamefont {S.}~\bibnamefont
  {Chun}}, \bibinfo {author} {\bibfnamefont {Y.}~\bibnamefont {Choi}}, \ and\
  \bibinfo {author} {\bibfnamefont {W.}~\bibnamefont {Park}},\ }\href@noop {}
  {\bibfield  {journal} {\bibinfo  {journal} {Carbon}\ }\textbf {\bibinfo
  {volume} {116}},\ \bibinfo {pages} {753} (\bibinfo {year}
  {2017})}\BibitemShut {NoStop}%
\bibitem [{\citenamefont {Fujita}\ \emph {et~al.}(2010)\citenamefont {Fujita},
  \citenamefont {Jalil},\ and\ \citenamefont {Tan}}]{Fujita10}%
  \BibitemOpen
  \bibfield  {author} {\bibinfo {author} {\bibfnamefont {T.}~\bibnamefont
  {Fujita}}, \bibinfo {author} {\bibfnamefont {M.~B.~A.}\ \bibnamefont
  {Jalil}}, \ and\ \bibinfo {author} {\bibfnamefont {S.~G.}\ \bibnamefont
  {Tan}},\ }\href@noop {} {\bibfield  {journal} {\bibinfo  {journal} {Appl.
  Phys. Lett.}\ }\textbf {\bibinfo {volume} {97}},\ \bibinfo {pages} {043508}
  (\bibinfo {year} {2010})}\BibitemShut {NoStop}%
\bibitem [{\citenamefont {Yesilyurt}\ \emph {et~al.}(2016)\citenamefont
  {Yesilyurt}, \citenamefont {Tan}, \citenamefont {Liang},\ and\ \citenamefont
  {Jalil}}]{Yesilyurt16}%
  \BibitemOpen
  \bibfield  {author} {\bibinfo {author} {\bibfnamefont {C.}~\bibnamefont
  {Yesilyurt}}, \bibinfo {author} {\bibfnamefont {S.~G.}\ \bibnamefont {Tan}},
  \bibinfo {author} {\bibfnamefont {G.~C.}\ \bibnamefont {Liang}}, \ and\
  \bibinfo {author} {\bibfnamefont {M.~B.~A.}\ \bibnamefont {Jalil}},\
  }\href@noop {} {\bibfield  {journal} {\bibinfo  {journal} {AIP Adv.}\
  }\textbf {\bibinfo {volume} {6}},\ \bibinfo {pages} {056303} (\bibinfo {year}
  {2016})}\BibitemShut {NoStop}%
\bibitem [{\citenamefont {Zhu}\ \emph {et~al.}(2015)\citenamefont {Zhu},
  \citenamefont {Stroscio},\ and\ \citenamefont {Li}}]{Zhu15}%
  \BibitemOpen
  \bibfield  {author} {\bibinfo {author} {\bibfnamefont {S.}~\bibnamefont
  {Zhu}}, \bibinfo {author} {\bibfnamefont {J.~A.}\ \bibnamefont {Stroscio}}, \
  and\ \bibinfo {author} {\bibfnamefont {T.}~\bibnamefont {Li}},\ }\href@noop
  {} {\bibfield  {journal} {\bibinfo  {journal} {Phys. Rev. Lett.}\ }\textbf
  {\bibinfo {volume} {115}},\ \bibinfo {pages} {245501} (\bibinfo {year}
  {2015})}\BibitemShut {NoStop}%
\bibitem [{\citenamefont {Jiang}\ \emph {et~al.}(2017)\citenamefont {Jiang},
  \citenamefont {Mao}, \citenamefont {Duan}, \citenamefont {Lai}, \citenamefont
  {Watanabe}, \citenamefont {Taniguchi},\ and\ \citenamefont
  {Andrei}}]{Jiang17}%
  \BibitemOpen
  \bibfield  {author} {\bibinfo {author} {\bibfnamefont {Y.}~\bibnamefont
  {Jiang}}, \bibinfo {author} {\bibfnamefont {J.}~\bibnamefont {Mao}}, \bibinfo
  {author} {\bibfnamefont {J.}~\bibnamefont {Duan}}, \bibinfo {author}
  {\bibfnamefont {X.}~\bibnamefont {Lai}}, \bibinfo {author} {\bibfnamefont
  {K.}~\bibnamefont {Watanabe}}, \bibinfo {author} {\bibfnamefont
  {T.}~\bibnamefont {Taniguchi}}, \ and\ \bibinfo {author} {\bibfnamefont
  {E.~Y.}\ \bibnamefont {Andrei}},\ }\href@noop {} {\bibfield  {journal}
  {\bibinfo  {journal} {Nano Lett.}\ }\textbf {\bibinfo {volume} {17}},\
  \bibinfo {pages} {2839} (\bibinfo {year} {2017})}\BibitemShut {NoStop}%
\bibitem [{\citenamefont {Guassi}\ \emph {et~al.}(2015)\citenamefont {Guassi},
  \citenamefont {Diniz}, \citenamefont {Sandler},\ and\ \citenamefont
  {Qu}}]{Guassi15}%
  \BibitemOpen
  \bibfield  {author} {\bibinfo {author} {\bibfnamefont {M.~R.}\ \bibnamefont
  {Guassi}}, \bibinfo {author} {\bibfnamefont {G.~S.}\ \bibnamefont {Diniz}},
  \bibinfo {author} {\bibfnamefont {N.}~\bibnamefont {Sandler}}, \ and\
  \bibinfo {author} {\bibfnamefont {F.~Y.}\ \bibnamefont {Qu}},\ }\href@noop {}
  {\bibfield  {journal} {\bibinfo  {journal} {Phys. Rev. B}\ }\textbf {\bibinfo
  {volume} {92}},\ \bibinfo {pages} {075426} (\bibinfo {year}
  {2015})}\BibitemShut {NoStop}%
\bibitem [{\citenamefont {Kretinin}\ \emph {et~al.}(2013)\citenamefont
  {Kretinin}, \citenamefont {Yu}, \citenamefont {Jalil}, \citenamefont {Cao},
  \citenamefont {Withers}, \citenamefont {Mishchenko}, \citenamefont
  {Katsnelson}, \citenamefont {Novoselov}, \citenamefont {Geim},\ and\
  \citenamefont {Guinea}}]{Kretinin13}%
  \BibitemOpen
  \bibfield  {author} {\bibinfo {author} {\bibfnamefont {A.}~\bibnamefont
  {Kretinin}}, \bibinfo {author} {\bibfnamefont {G.~L.}\ \bibnamefont {Yu}},
  \bibinfo {author} {\bibfnamefont {R.}~\bibnamefont {Jalil}}, \bibinfo
  {author} {\bibfnamefont {Y.}~\bibnamefont {Cao}}, \bibinfo {author}
  {\bibfnamefont {F.}~\bibnamefont {Withers}}, \bibinfo {author} {\bibfnamefont
  {A.}~\bibnamefont {Mishchenko}}, \bibinfo {author} {\bibfnamefont {M.~I.}\
  \bibnamefont {Katsnelson}}, \bibinfo {author} {\bibfnamefont {K.~S.}\
  \bibnamefont {Novoselov}}, \bibinfo {author} {\bibfnamefont {A.~K.}\
  \bibnamefont {Geim}}, \ and\ \bibinfo {author} {\bibfnamefont
  {F.}~\bibnamefont {Guinea}},\ }\href@noop {} {\bibfield  {journal} {\bibinfo
  {journal} {Phys. Rev. B}\ }\textbf {\bibinfo {volume} {88}},\ \bibinfo
  {pages} {165427} (\bibinfo {year} {2013})}\BibitemShut {NoStop}%
\bibitem [{\citenamefont {Chaves}\ \emph {et~al.}(2014)\citenamefont {Chaves},
  \citenamefont {Jimenez}, \citenamefont {Cummings},\ and\ \citenamefont
  {Roche}}]{Chaves14}%
  \BibitemOpen
  \bibfield  {author} {\bibinfo {author} {\bibfnamefont {F.~A.}\ \bibnamefont
  {Chaves}}, \bibinfo {author} {\bibfnamefont {D.}~\bibnamefont {Jimenez}},
  \bibinfo {author} {\bibfnamefont {A.~W.}\ \bibnamefont {Cummings}}, \ and\
  \bibinfo {author} {\bibfnamefont {S.}~\bibnamefont {Roche}},\ }\href@noop {}
  {\bibfield  {journal} {\bibinfo  {journal} {J. Appl. Phys.}\ }\textbf
  {\bibinfo {volume} {115}},\ \bibinfo {pages} {164513} (\bibinfo {year}
  {2014})}\BibitemShut {NoStop}%
\bibitem [{\citenamefont {Tworzydlo}\ \emph {et~al.}(2006)\citenamefont
  {Tworzydlo}, \citenamefont {Trauzettel}, \citenamefont {Titov}, \citenamefont
  {Rycerz},\ and\ \citenamefont {Beenakker}}]{Tworzydlo06}%
  \BibitemOpen
  \bibfield  {author} {\bibinfo {author} {\bibfnamefont {J.}~\bibnamefont
  {Tworzydlo}}, \bibinfo {author} {\bibfnamefont {B.}~\bibnamefont
  {Trauzettel}}, \bibinfo {author} {\bibfnamefont {M.}~\bibnamefont {Titov}},
  \bibinfo {author} {\bibfnamefont {A.}~\bibnamefont {Rycerz}}, \ and\ \bibinfo
  {author} {\bibfnamefont {C.~W.}\ \bibnamefont {Beenakker}},\ }\href@noop {}
  {\bibfield  {journal} {\bibinfo  {journal} {Phys. Rev. Lett.}\ }\textbf
  {\bibinfo {volume} {96}},\ \bibinfo {pages} {246802} (\bibinfo {year}
  {2006})}\BibitemShut {NoStop}%
\bibitem [{\citenamefont {Yigen}\ and\ \citenamefont
  {Champagne}(2014)}]{Yigen14}%
  \BibitemOpen
  \bibfield  {author} {\bibinfo {author} {\bibfnamefont {S.}~\bibnamefont
  {Yigen}}\ and\ \bibinfo {author} {\bibfnamefont {A.~R.}\ \bibnamefont
  {Champagne}},\ }\href@noop {} {\bibfield  {journal} {\bibinfo  {journal}
  {Nano Lett.}\ }\textbf {\bibinfo {volume} {14}},\ \bibinfo {pages} {289}
  (\bibinfo {year} {2014})}\BibitemShut {NoStop}%
\bibitem [{\citenamefont {Wu}\ \emph {et~al.}(2012)\citenamefont {Wu},
  \citenamefont {Perebeinos}, \citenamefont {Lin}, \citenamefont {Low},
  \citenamefont {Xia},\ and\ \citenamefont {Avouris}}]{Wu12}%
  \BibitemOpen
  \bibfield  {author} {\bibinfo {author} {\bibfnamefont {Y.~Q.}\ \bibnamefont
  {Wu}}, \bibinfo {author} {\bibfnamefont {V.}~\bibnamefont {Perebeinos}},
  \bibinfo {author} {\bibfnamefont {Y.~M.}\ \bibnamefont {Lin}}, \bibinfo
  {author} {\bibfnamefont {T.}~\bibnamefont {Low}}, \bibinfo {author}
  {\bibfnamefont {F.~N.}\ \bibnamefont {Xia}}, \ and\ \bibinfo {author}
  {\bibfnamefont {P.}~\bibnamefont {Avouris}},\ }\href@noop {} {\bibfield
  {journal} {\bibinfo  {journal} {Nano Lett.}\ }\textbf {\bibinfo {volume}
  {12}},\ \bibinfo {pages} {1417} (\bibinfo {year} {2012})}\BibitemShut
  {NoStop}%
\bibitem [{\citenamefont {McRae}\ \emph {et~al.}(2017)\citenamefont {McRae},
  \citenamefont {Tayari}, \citenamefont {Porter},\ and\ \citenamefont
  {Champagne}}]{McRae17}%
  \BibitemOpen
  \bibfield  {author} {\bibinfo {author} {\bibfnamefont {A.~C.}\ \bibnamefont
  {McRae}}, \bibinfo {author} {\bibfnamefont {V.}~\bibnamefont {Tayari}},
  \bibinfo {author} {\bibfnamefont {J.~M.}\ \bibnamefont {Porter}}, \ and\
  \bibinfo {author} {\bibfnamefont {A.~R.}\ \bibnamefont {Champagne}},\
  }\href@noop {} {\bibfield  {journal} {\bibinfo  {journal} {Nat. Commun.}\
  }\textbf {\bibinfo {volume} {8}},\ \bibinfo {pages} {15491} (\bibinfo {year}
  {2017})}\BibitemShut {NoStop}%
\bibitem [{\citenamefont {Huang}\ \emph {et~al.}(2009)\citenamefont {Huang},
  \citenamefont {Yan}, \citenamefont {Chen}, \citenamefont {Song},
  \citenamefont {Heinz},\ and\ \citenamefont {Hone}}]{Huang09}%
  \BibitemOpen
  \bibfield  {author} {\bibinfo {author} {\bibfnamefont {M.~Y.}\ \bibnamefont
  {Huang}}, \bibinfo {author} {\bibfnamefont {H.~G.}\ \bibnamefont {Yan}},
  \bibinfo {author} {\bibfnamefont {C.~Y.}\ \bibnamefont {Chen}}, \bibinfo
  {author} {\bibfnamefont {D.~H.}\ \bibnamefont {Song}}, \bibinfo {author}
  {\bibfnamefont {T.~F.}\ \bibnamefont {Heinz}}, \ and\ \bibinfo {author}
  {\bibfnamefont {J.}~\bibnamefont {Hone}},\ }\href@noop {} {\bibfield
  {journal} {\bibinfo  {journal} {P. Natl. Acad. Sci. USA}\ }\textbf {\bibinfo
  {volume} {106}},\ \bibinfo {pages} {7304} (\bibinfo {year}
  {2009})}\BibitemShut {NoStop}%
\bibitem [{\citenamefont {Andrei}\ \emph {et~al.}(2012)\citenamefont {Andrei},
  \citenamefont {Li},\ and\ \citenamefont {Du}}]{Andrei12}%
  \BibitemOpen
  \bibfield  {author} {\bibinfo {author} {\bibfnamefont {E.~Y.}\ \bibnamefont
  {Andrei}}, \bibinfo {author} {\bibfnamefont {G.}~\bibnamefont {Li}}, \ and\
  \bibinfo {author} {\bibfnamefont {X.}~\bibnamefont {Du}},\ }\href@noop {}
  {\bibfield  {journal} {\bibinfo  {journal} {Rep. Prog. Phys.}\ }\textbf
  {\bibinfo {volume} {75}},\ \bibinfo {pages} {056501} (\bibinfo {year}
  {2012})}\BibitemShut {NoStop}%
\bibitem [{\citenamefont {Bolotin}\ \emph {et~al.}(2008)\citenamefont
  {Bolotin}, \citenamefont {Sikes}, \citenamefont {Jiang}, \citenamefont
  {Klima}, \citenamefont {Fudenberg}, \citenamefont {Hone}, \citenamefont
  {Kim},\ and\ \citenamefont {Stormer}}]{Bolotin08}%
  \BibitemOpen
  \bibfield  {author} {\bibinfo {author} {\bibfnamefont {K.~I.}\ \bibnamefont
  {Bolotin}}, \bibinfo {author} {\bibfnamefont {K.~J.}\ \bibnamefont {Sikes}},
  \bibinfo {author} {\bibfnamefont {Z.}~\bibnamefont {Jiang}}, \bibinfo
  {author} {\bibfnamefont {M.}~\bibnamefont {Klima}}, \bibinfo {author}
  {\bibfnamefont {G.}~\bibnamefont {Fudenberg}}, \bibinfo {author}
  {\bibfnamefont {J.}~\bibnamefont {Hone}}, \bibinfo {author} {\bibfnamefont
  {P.}~\bibnamefont {Kim}}, \ and\ \bibinfo {author} {\bibfnamefont {H.~L.}\
  \bibnamefont {Stormer}},\ }\href@noop {} {\bibfield  {journal} {\bibinfo
  {journal} {Solid State Commun.}\ }\textbf {\bibinfo {volume} {146}},\
  \bibinfo {pages} {351} (\bibinfo {year} {2008})}\BibitemShut {NoStop}%
\bibitem [{\citenamefont {Tayari}\ \emph {et~al.}(2015)\citenamefont {Tayari},
  \citenamefont {McRae}, \citenamefont {Yigen}, \citenamefont {Island},
  \citenamefont {Porter},\ and\ \citenamefont {Champagne}}]{Tayari15}%
  \BibitemOpen
  \bibfield  {author} {\bibinfo {author} {\bibfnamefont {V.}~\bibnamefont
  {Tayari}}, \bibinfo {author} {\bibfnamefont {A.~C.}\ \bibnamefont {McRae}},
  \bibinfo {author} {\bibfnamefont {S.}~\bibnamefont {Yigen}}, \bibinfo
  {author} {\bibfnamefont {J.~O.}\ \bibnamefont {Island}}, \bibinfo {author}
  {\bibfnamefont {J.~M.}\ \bibnamefont {Porter}}, \ and\ \bibinfo {author}
  {\bibfnamefont {A.~R.}\ \bibnamefont {Champagne}},\ }\href@noop {} {\bibfield
   {journal} {\bibinfo  {journal} {Nano Lett.}\ }\textbf {\bibinfo {volume}
  {15}},\ \bibinfo {pages} {114} (\bibinfo {year} {2015})}\BibitemShut
  {NoStop}%
\bibitem [{\citenamefont {Champagne}\ \emph {et~al.}(2005)\citenamefont
  {Champagne}, \citenamefont {Pasupathy},\ and\ \citenamefont
  {Ralph}}]{Champagne05}%
  \BibitemOpen
  \bibfield  {author} {\bibinfo {author} {\bibfnamefont {A.~R.}\ \bibnamefont
  {Champagne}}, \bibinfo {author} {\bibfnamefont {A.~N.}\ \bibnamefont
  {Pasupathy}}, \ and\ \bibinfo {author} {\bibfnamefont {D.~C.}\ \bibnamefont
  {Ralph}},\ }\href@noop {} {\bibfield  {journal} {\bibinfo  {journal} {Nano
  Lett.}\ }\textbf {\bibinfo {volume} {5}},\ \bibinfo {pages} {305} (\bibinfo
  {year} {2005})}\BibitemShut {NoStop}%
\bibitem [{\citenamefont {Parks}\ \emph {et~al.}(2007)\citenamefont {Parks},
  \citenamefont {Champagne}, \citenamefont {Hutchison}, \citenamefont
  {Flores-Torres}, \citenamefont {Abruna},\ and\ \citenamefont
  {Ralph}}]{Parks07}%
  \BibitemOpen
  \bibfield  {author} {\bibinfo {author} {\bibfnamefont {J.~J.}\ \bibnamefont
  {Parks}}, \bibinfo {author} {\bibfnamefont {A.~R.}\ \bibnamefont
  {Champagne}}, \bibinfo {author} {\bibfnamefont {G.~R.}\ \bibnamefont
  {Hutchison}}, \bibinfo {author} {\bibfnamefont {S.}~\bibnamefont
  {Flores-Torres}}, \bibinfo {author} {\bibfnamefont {H.~D.}\ \bibnamefont
  {Abruna}}, \ and\ \bibinfo {author} {\bibfnamefont {D.~C.}\ \bibnamefont
  {Ralph}},\ }\href@noop {} {\bibfield  {journal} {\bibinfo  {journal} {Phys.
  Rev. Lett.}\ }\textbf {\bibinfo {volume} {99}},\ \bibinfo {pages} {026601}
  (\bibinfo {year} {2007})}\BibitemShut {NoStop}%
\bibitem [{\citenamefont {Parks}\ \emph {et~al.}(2010)\citenamefont {Parks},
  \citenamefont {Champagne}, \citenamefont {Costi}, \citenamefont {Shum},
  \citenamefont {Pasupathy}, \citenamefont {Neuscamman}, \citenamefont
  {Flores-Torres}, \citenamefont {Cornaglia}, \citenamefont {Aligia},
  \citenamefont {Balseiro}, \citenamefont {Chan}, \citenamefont {Abruna},\ and\
  \citenamefont {Ralph}}]{Parks10}%
  \BibitemOpen
  \bibfield  {author} {\bibinfo {author} {\bibfnamefont {J.~J.}\ \bibnamefont
  {Parks}}, \bibinfo {author} {\bibfnamefont {A.~R.}\ \bibnamefont
  {Champagne}}, \bibinfo {author} {\bibfnamefont {T.~A.}\ \bibnamefont
  {Costi}}, \bibinfo {author} {\bibfnamefont {W.~W.}\ \bibnamefont {Shum}},
  \bibinfo {author} {\bibfnamefont {A.~N.}\ \bibnamefont {Pasupathy}}, \bibinfo
  {author} {\bibfnamefont {E.}~\bibnamefont {Neuscamman}}, \bibinfo {author}
  {\bibfnamefont {S.}~\bibnamefont {Flores-Torres}}, \bibinfo {author}
  {\bibfnamefont {P.~S.}\ \bibnamefont {Cornaglia}}, \bibinfo {author}
  {\bibfnamefont {A.~A.}\ \bibnamefont {Aligia}}, \bibinfo {author}
  {\bibfnamefont {C.~A.}\ \bibnamefont {Balseiro}}, \bibinfo {author}
  {\bibfnamefont {G.~K.~L.}\ \bibnamefont {Chan}}, \bibinfo {author}
  {\bibfnamefont {H.~D.}\ \bibnamefont {Abruna}}, \ and\ \bibinfo {author}
  {\bibfnamefont {D.~C.}\ \bibnamefont {Ralph}},\ }\href@noop {} {\bibfield
  {journal} {\bibinfo  {journal} {Science}\ }\textbf {\bibinfo {volume}
  {328}},\ \bibinfo {pages} {1370} (\bibinfo {year} {2010})}\BibitemShut
  {NoStop}%
\bibitem [{\citenamefont {Goldsche}\ \emph {et~al.}(2018)\citenamefont
  {Goldsche}, \citenamefont {Sonntag}, \citenamefont {Khodkov}, \citenamefont
  {Verbiest}, \citenamefont {Reichardt}, \citenamefont {Neumann}, \citenamefont
  {Ouaj}, \citenamefont {von~den Driesch}, \citenamefont {Buca},\ and\
  \citenamefont {Stampfer}}]{Goldsche18}%
  \BibitemOpen
  \bibfield  {author} {\bibinfo {author} {\bibfnamefont {M.}~\bibnamefont
  {Goldsche}}, \bibinfo {author} {\bibfnamefont {J.}~\bibnamefont {Sonntag}},
  \bibinfo {author} {\bibfnamefont {T.}~\bibnamefont {Khodkov}}, \bibinfo
  {author} {\bibfnamefont {G.~J.}\ \bibnamefont {Verbiest}}, \bibinfo {author}
  {\bibfnamefont {S.}~\bibnamefont {Reichardt}}, \bibinfo {author}
  {\bibfnamefont {C.}~\bibnamefont {Neumann}}, \bibinfo {author} {\bibfnamefont
  {T.}~\bibnamefont {Ouaj}}, \bibinfo {author} {\bibfnamefont {N.}~\bibnamefont
  {von~den Driesch}}, \bibinfo {author} {\bibfnamefont {D.}~\bibnamefont
  {Buca}}, \ and\ \bibinfo {author} {\bibfnamefont {C.}~\bibnamefont
  {Stampfer}},\ }\href@noop {} {\bibfield  {journal} {\bibinfo  {journal} {Nano
  Lett.}\ }\textbf {\bibinfo {volume} {18}},\ \bibinfo {pages} {1707} (\bibinfo
  {year} {2018})}\BibitemShut {NoStop}%
\bibitem [{\citenamefont {Sarwat}\ \emph {et~al.}(2018)\citenamefont {Sarwat},
  \citenamefont {Tweedie}, \citenamefont {Porter}, \citenamefont {Zhou},
  \citenamefont {Sheng}, \citenamefont {Mol}, \citenamefont {Warner},\ and\
  \citenamefont {Bhaskaran}}]{Sarwat18}%
  \BibitemOpen
  \bibfield  {author} {\bibinfo {author} {\bibfnamefont {S.~G.}\ \bibnamefont
  {Sarwat}}, \bibinfo {author} {\bibfnamefont {M.}~\bibnamefont {Tweedie}},
  \bibinfo {author} {\bibfnamefont {B.~F.}\ \bibnamefont {Porter}}, \bibinfo
  {author} {\bibfnamefont {Y.}~\bibnamefont {Zhou}}, \bibinfo {author}
  {\bibfnamefont {Y.}~\bibnamefont {Sheng}}, \bibinfo {author} {\bibfnamefont
  {J.}~\bibnamefont {Mol}}, \bibinfo {author} {\bibfnamefont {J.}~\bibnamefont
  {Warner}}, \ and\ \bibinfo {author} {\bibfnamefont {H.}~\bibnamefont
  {Bhaskaran}},\ }\href@noop {} {\bibfield  {journal} {\bibinfo  {journal}
  {Nano Lett.}\ }\textbf {\bibinfo {volume} {18}},\ \bibinfo {pages} {2467}
  (\bibinfo {year} {2018})}\BibitemShut {NoStop}%
\bibitem [{\citenamefont {Nix}\ and\ \citenamefont {MacNair}(1941)}]{Nix41}%
  \BibitemOpen
  \bibfield  {author} {\bibinfo {author} {\bibfnamefont {F.~C.}\ \bibnamefont
  {Nix}}\ and\ \bibinfo {author} {\bibfnamefont {D.}~\bibnamefont {MacNair}},\
  }\href@noop {} {\bibfield  {journal} {\bibinfo  {journal} {Phys. Rev.}\
  }\textbf {\bibinfo {volume} {60}},\ \bibinfo {pages} {597} (\bibinfo {year}
  {1941})}\BibitemShut {NoStop}%
\bibitem [{\citenamefont {Sundaram}\ \emph {et~al.}(2011)\citenamefont
  {Sundaram}, \citenamefont {Steiner}, \citenamefont {Chiu}, \citenamefont
  {Engel}, \citenamefont {Bol}, \citenamefont {Krupke}, \citenamefont
  {Burghard}, \citenamefont {Kern},\ and\ \citenamefont
  {Avouris}}]{Sundaram11}%
  \BibitemOpen
  \bibfield  {author} {\bibinfo {author} {\bibfnamefont {R.~S.}\ \bibnamefont
  {Sundaram}}, \bibinfo {author} {\bibfnamefont {M.}~\bibnamefont {Steiner}},
  \bibinfo {author} {\bibfnamefont {H.~Y.}\ \bibnamefont {Chiu}}, \bibinfo
  {author} {\bibfnamefont {M.}~\bibnamefont {Engel}}, \bibinfo {author}
  {\bibfnamefont {A.~A.}\ \bibnamefont {Bol}}, \bibinfo {author} {\bibfnamefont
  {R.}~\bibnamefont {Krupke}}, \bibinfo {author} {\bibfnamefont
  {M.}~\bibnamefont {Burghard}}, \bibinfo {author} {\bibfnamefont
  {K.}~\bibnamefont {Kern}}, \ and\ \bibinfo {author} {\bibfnamefont
  {P.}~\bibnamefont {Avouris}},\ }\href@noop {} {\bibfield  {journal} {\bibinfo
   {journal} {Nano Lett.}\ }\textbf {\bibinfo {volume} {11}},\ \bibinfo {pages}
  {3833} (\bibinfo {year} {2011})}\BibitemShut {NoStop}%
\bibitem [{\citenamefont {Heinze}\ \emph {et~al.}(2002)\citenamefont {Heinze},
  \citenamefont {Tersoff}, \citenamefont {Martel}, \citenamefont {Derycke},
  \citenamefont {Appenzeller},\ and\ \citenamefont {Avouris}}]{Heinze02}%
  \BibitemOpen
  \bibfield  {author} {\bibinfo {author} {\bibfnamefont {S.}~\bibnamefont
  {Heinze}}, \bibinfo {author} {\bibfnamefont {J.}~\bibnamefont {Tersoff}},
  \bibinfo {author} {\bibfnamefont {R.}~\bibnamefont {Martel}}, \bibinfo
  {author} {\bibfnamefont {V.}~\bibnamefont {Derycke}}, \bibinfo {author}
  {\bibfnamefont {J.}~\bibnamefont {Appenzeller}}, \ and\ \bibinfo {author}
  {\bibfnamefont {P.}~\bibnamefont {Avouris}},\ }\href@noop {} {\bibfield
  {journal} {\bibinfo  {journal} {Phys. Rev. Lett.}\ }\textbf {\bibinfo
  {volume} {89}},\ \bibinfo {pages} {106801} (\bibinfo {year}
  {2002})}\BibitemShut {NoStop}%
\bibitem [{\citenamefont {Pereira}\ \emph {et~al.}(2009)\citenamefont
  {Pereira}, \citenamefont {Castro~Neto},\ and\ \citenamefont
  {Peres}}]{Pereira09}%
  \BibitemOpen
  \bibfield  {author} {\bibinfo {author} {\bibfnamefont {V.~M.}\ \bibnamefont
  {Pereira}}, \bibinfo {author} {\bibfnamefont {A.~H.}\ \bibnamefont
  {Castro~Neto}}, \ and\ \bibinfo {author} {\bibfnamefont {N.~M.~R.}\
  \bibnamefont {Peres}},\ }\href@noop {} {\bibfield  {journal} {\bibinfo
  {journal} {Phys. Rev. B}\ }\textbf {\bibinfo {volume} {80}},\ \bibinfo
  {pages} {045401} (\bibinfo {year} {2009})}\BibitemShut {NoStop}%
\bibitem [{\citenamefont {Choi}\ \emph {et~al.}(2010)\citenamefont {Choi},
  \citenamefont {Jhi},\ and\ \citenamefont {Son}}]{Choi10}%
  \BibitemOpen
  \bibfield  {author} {\bibinfo {author} {\bibfnamefont {S.-M.}\ \bibnamefont
  {Choi}}, \bibinfo {author} {\bibfnamefont {S.-H.}\ \bibnamefont {Jhi}}, \
  and\ \bibinfo {author} {\bibfnamefont {Y.-W.}\ \bibnamefont {Son}},\
  }\href@noop {} {\bibfield  {journal} {\bibinfo  {journal} {Phys. Rev. B}\
  }\textbf {\bibinfo {volume} {81}},\ \bibinfo {pages} {081407} (\bibinfo
  {year} {2010})}\BibitemShut {NoStop}%
\bibitem [{\citenamefont {Rickhaus}\ \emph {et~al.}(2013)\citenamefont
  {Rickhaus}, \citenamefont {Maurand}, \citenamefont {Liu}, \citenamefont
  {Weiss}, \citenamefont {Richter},\ and\ \citenamefont
  {Schonenberger}}]{Rickhaus13}%
  \BibitemOpen
  \bibfield  {author} {\bibinfo {author} {\bibfnamefont {P.}~\bibnamefont
  {Rickhaus}}, \bibinfo {author} {\bibfnamefont {R.}~\bibnamefont {Maurand}},
  \bibinfo {author} {\bibfnamefont {M.~H.}\ \bibnamefont {Liu}}, \bibinfo
  {author} {\bibfnamefont {M.}~\bibnamefont {Weiss}}, \bibinfo {author}
  {\bibfnamefont {K.}~\bibnamefont {Richter}}, \ and\ \bibinfo {author}
  {\bibfnamefont {C.}~\bibnamefont {Schonenberger}},\ }\href@noop {} {\bibfield
   {journal} {\bibinfo  {journal} {Nat. Commun.}\ }\textbf {\bibinfo {volume}
  {4}},\ \bibinfo {pages} {2342} (\bibinfo {year} {2013})}\BibitemShut
  {NoStop}%
\bibitem [{\citenamefont {Anzi}\ \emph {et~al.}(2018)\citenamefont {Anzi},
  \citenamefont {Mansouri}, \citenamefont {Pedrinazzi}, \citenamefont
  {Guerriero}, \citenamefont {Fiocco}, \citenamefont {Pesquera}, \citenamefont
  {Centeno}, \citenamefont {Zurutuza}, \citenamefont {Behnam}, \citenamefont
  {Carrion}, \citenamefont {Pop},\ and\ \citenamefont {Sordan}}]{Anzi18}%
  \BibitemOpen
  \bibfield  {author} {\bibinfo {author} {\bibfnamefont {L.}~\bibnamefont
  {Anzi}}, \bibinfo {author} {\bibfnamefont {A.}~\bibnamefont {Mansouri}},
  \bibinfo {author} {\bibfnamefont {P.}~\bibnamefont {Pedrinazzi}}, \bibinfo
  {author} {\bibfnamefont {E.}~\bibnamefont {Guerriero}}, \bibinfo {author}
  {\bibfnamefont {M.}~\bibnamefont {Fiocco}}, \bibinfo {author} {\bibfnamefont
  {A.}~\bibnamefont {Pesquera}}, \bibinfo {author} {\bibfnamefont
  {A.}~\bibnamefont {Centeno}}, \bibinfo {author} {\bibfnamefont
  {A.}~\bibnamefont {Zurutuza}}, \bibinfo {author} {\bibfnamefont
  {A.}~\bibnamefont {Behnam}}, \bibinfo {author} {\bibfnamefont {E.~A.}\
  \bibnamefont {Carrion}}, \bibinfo {author} {\bibfnamefont {E.}~\bibnamefont
  {Pop}}, \ and\ \bibinfo {author} {\bibfnamefont {R.}~\bibnamefont {Sordan}},\
  }\href@noop {} {\bibfield  {journal} {\bibinfo  {journal} {2D Mater.}\
  }\textbf {\bibinfo {volume} {5}},\ \bibinfo {pages} {025014} (\bibinfo {year}
  {2018})}\BibitemShut {NoStop}%
\end{thebibliography}
\end{document}